\documentclass[review, sort&compress]{elsarticle}

\usepackage{lineno,hyperref}

\usepackage{amsmath,amssymb,mathtools}
\usepackage[scientific-notation=true]{siunitx}
\usepackage{graphicx}
\usepackage{pgfplots}
\usepackage{fullpage}
\usepackage{changepage}
\usepackage{caption,subcaption}
\usepackage{anyfontsize}
\usepackage{booktabs}
\usepackage{tabu}
\usepackage{multirow,dcolumn,longtable,adjustbox}
\usepackage{lscape}
\usepackage{enumerate}
\usepackage{enumitem}
\usepackage{float}
\usepackage{hyperref}
\usepackage{xcolor,soul}
\usepackage[capitalise]{cleveref}
\usepackage{comment}

\definecolor{mygray}{RGB}{127,127,127}
\definecolor{myred}{RGB}{177,0,0}
\definecolor{mygreen}{RGB}{0,177,0}
\definecolor{myblue}{RGB}{0,0,177}

\newcommand{\One}{Al$_{16}$Cr$_{25}$Fe$_{35}$Ni$_{20}$Ti$_4$~}
\newcommand{\Two}{Al$_{20}$Cr$_{20}$Fe$_{35}$Ni$_{20}$Ti$_5$~}
\newcommand{\Three}{Al$_{24}$Cr$_{15}$Fe$_{35}$Ni$_{20}$Ti$_6$~}

\newcolumntype{d}{D{.}{.}{-1} }

\modulolinenumbers[5]

\journal{Acta Materialia}

\begin{document}

\begin{frontmatter}

\title{Towards Superior High Temperature Properties in Low Density AlCrFeNiTi Compositionally Complex Alloys}

\author[max-planck]{Silas Wolff-Goodrich\corref{corr_author}}
\cortext[corr_author]{Corresponding author}
\ead{s.wolffgoodrich@mpie.de}
\author[bayreuth]{Sebastian Haas}
\author[bayreuth]{Uwe Glatzel}
\author[max-planck]{Christian H. Liebscher}

\address[max-planck]{Max-Planck-Institut f{\"u}r Eisenforschung GmbH, Max-Planck-Stra{\ss}e 1, 40237 D{\"u}sseldorf}
\address[bayreuth]{Universit{\"a}t Bayreuth, Lehrstuhl Metallische Werkstoffe, TAO-Geb{\"a}ude, Prof.-R{\"u}diger-Bormann-Str. 1, 95447 Bayreuth}

\begin{abstract}

Three novel precipitation strengthened bcc alloys which exhibit a smooth microstructural gradient with composition have been fabricated in bulk form by induction casting. All three alloys are comprised of a mixture of disordered A2-(Fe, Cr) and L2$_1$-ordered (Ni, Fe)$_{2}$AlTi type phases both as-cast and after long-term annealing at 900 $^{\circ}$C. The ratio of disordered to ordered phase, primary dendrite fraction, and overall microstructural coarseness all decrease as Cr is replaced by Al and Ti. Differences in phase composition are quantified through domain averaged principal component analysis of energy dispersive spectroscopy data obtained during scanning transmission electron microscopy. Bulk tensile testing reveals retained strengths of nearly 250 MPa up to 900 $^{\circ}$C for the alloys which contain a nanoscale maze-like arrangement of ordered and disordered phases. One alloy, containing a duplex microstructure with ductile dendritic regions and highly creep resistant interdendritic regions, shows a promising balance between high temperature ductility and strength. For this alloy, tension creep testing was carried out at 700, 750, and 800 $^{\circ}$C for a broad range of loading conditions and revealed upper bound creep rates which surpass similar ferritic superalloys and rival those of several conventionally employed high temperature structural alloys, including Inconel 617 and 718, at much lower density and raw material cost.\\

\end{abstract}


\end{frontmatter}


\section{Introduction}

The high entropy alloy (HEA) and compositionally complex alloy (CCA) design approaches have in recent years yielded many promising new alloys for high temperature structural applications. Numerous studies have been published on fcc-structured CCAs strengthened by L1$_2$ precipitates, analogous to Ni-base superalloys, which show impressive high temperature mechanical properties, in both compression and tension~\cite{Tsao2017, Haas2019, Chen2020, Zhang2020}. While fcc-structured CCAs typically have superior ductility, several bcc-structured CCAs have shown promising high temperature strengths with reasonable ductility. Lim et al. studied the high temperature compression behaviour of the equiatomic AlCoCrFeNi alloy and found a yield strength of 201 MPa with over 50\% plastic strain at 900 $^{\circ}$C. This alloy was found to be composed of nearly equal parts of dendritic regions, with a B2 matrix and A2 structured cuboidal precipitates, and interdendritic regions, with interpenetrating A2 and B2 domains
~\cite{Lim2017}. Stepanov et al. explored a series of alloy compositions in the system Fe$_{36}$X$_{21}$Cr$_{18}$Ni$_{15}$Al$_{10}$ and Fe$_{35}$X$_{20}$Cr$_{17}$Ni$_{12}$Al$_{12}$Ti$_4$ with X=(Co, Mn), and found a compressive yield stress of 285 MPa and over 50\% strain-to-failure at 800 $^{\circ}$C for the Fe$_{35}$Co$_{20}$Cr$_{17}$Ni$_{12}$Al$_{12}$Ti$_4$ composition, which contains a mixture of ordered and disordered bcc phases, with small amounts of disordered fcc phases near grain boundaries~\cite{Stepanov2018}. In a related work, Shaysultanov et al. measured a tensile yield stress of 310 MPa and strain-to-failure of 55\% for the Fe$_{36}$Mn$_{21}$Cr$_{18}$Ni$_{15}$Al$_{10}$ alloy at 600 $^{\circ}$C
~\cite{Shaysultanov2017}. A compressive yield stress of nearly 400 MPa at 800 $^{\circ}$C was reported by Zhou et al. for the B2 strengthened bcc-structured Fe$_{34}$Cr$_{34}$Ni$_{14}$Al$_{14}$Co$_4$ alloy~\cite{Zhou2018}. In general, however, there is presently a lack of high temperature tensile test data for bcc-structured CCAs---let alone creep data---presumably due to the low tensile ductility that is often observed in such alloys.\\ 

The Al-Cr-Fe-Ni-Ti alloy system comprises several well studied binary and ternary systems. Of particular relevance to the present work are the Al-Cr-Fe, Al-Cr-Ni, Al-Fe-Ti, and Al-Ni-Ti systems. The phase and microstructure formation for large portions of the composition space for each of these systems is well understood, especially in regions of single and dual phase A2, B2, and/or L2$_1$ (Heusler phase) stability. B2-structured FeAl and D0$_3$-structured Fe$_3$Al as structural materials have the benefits of relatively low cost, low density, very good high temperature oxidation resistance, and moderate high temperature strength~\cite{Yoshimi1996, Morris2004}. Similarly, B2-structured NiAl and L1$_2$-structured Ni$_3$Al show the same benefits as FeAl, but have higher high-temperature strength and better creep resistance~\cite{Darolia1991, Rudy1986}. The addition of titanium to both the Al-Fe and Al-Ni systems allows formation of the L2$_1$-structured Heusler phase, as well as Laves phases. Single phase L2$_1$-Ni$_2$TiAl as well as dual phase B2-NiAl+L2$_1$-Ni$_2$TiAl ($\beta-\beta'$) have been shown to possess high temperature creep resistance that is directly comparable with high-performing $\gamma-\gamma'$ Ni-base superalloys up to around 1000 $^{\circ}$C~\cite{Darolia1991, Polvani1976, Strutt1976}. For applications with rotating components, where the maximum stress is proportional to the alloy density, these Heusler phase alloys may surpass the creep resistance of Ni-base superalloys.\\

The main issue with the intermetallic alloys mentioned above is their very low ductility and fracture toughness in polycrystalline form, even at elevated temperatures. One prominent class of alloys that was originally developed as a low cost alternative to Ni-base superalloys for structural applications up to approximately 760 $^{\circ}$C which do not suffer from this drawback are so called Fe-base superalloys. This class of alloys makes use of coherently precipitated B2-NiAl and/or L2$_1$-Ni$_2$TiAl as the primary strengthening phase(s) for an A2 Fe-rich matrix~\cite{Bhadeshia2001, Calderon1984, Calderon1988, Liebscher2013, Sun2013, Liebscher2015, Song2015, Song2017, Song2017a, Baik2018}. While certainly more ductile than the pure B2/L2$_1$ alloys, these alloys are inherently limited in their maximum application temperature by the rapid decrease in strength and creep resistance of the A2 matrix at temperatures above about 700 $^{\circ}$C. One way to maintain the high temperature mechanical properties is to increase the ordered phase content, however, above a certain amount of Ni and Al, there is no longer a clear matrix precipitate relationship, as the ordered phase domains become fully continuous~\cite{Stallybrass2004}. Such maze-like or lamellar microstructures are widely reported in the HEA/CCA literature
(\cite{Singh2011, Chen2017, Ma2018, Tian2019, Wolff-Goodrich2021} to list just a few) and, while apparently very strong, have been shown to be very brittle~\cite{Ma2018, Wolff-Goodrich2021}.\\ 

In a previous investigation, we screened the phase and microstructure formation across a large portion of the Al-Cr-Fe-Ni-Ti composition space~\cite{Wolff-Goodrich2021}. Microstructures consisting of a primary A2 matrix with equiaxed B2/L2$_1$ precipitates and interdendritic B2/L2$_1$ domains were observed, as well as microstructures consisting entirely of a fine maze-like arrangement of A2 and B2/L2$_1$ domains. One alloy, with composition \Two, was found to consist of primary dendrites containing the A2 matrix-B2/L2$_1$ precipitate morphology, and interdendritic regions containing the maze-like morphology, with the interdendritic regions occupying a slightly larger total portion of the structure. A similar microstructure has been observed in the work of Tian et al. for equiatomic AlCoCrFeNi, where the authors observed a room temperature compressive yield stress of about 1630 MPa with 16\% strain to failure~\cite{Tian2019}.\\

In the present work, we explore a set of low density CCAs where maximised contributions from precipitation hardening and solid solution strengthening combine for exceptional high temperature mechanical properties up to 800 $^{\circ}$C. We will illustrate the effects of adjusting the ratio of matrix-precipitate to maze-like morphology, as well as the phase compositions, on the high temperature performance of these alloys. We have tailored a microstructural gradient consisting of three compositions, centered at \One, where Al and Ti are substituted for Cr while holding Fe and Ni constant. The compositions of the three alloys to be studied can be designated as Al$_{4x}$Cr$_{5*(9-x)}$Fe$_{35}$Ni$_{20}$Ti$_{x}$ for $x\in\{4,5,6\}$ (in at.\%). The particular ratio of Al:Ti=4:1 was chosen based on findings that an optimum tension creep resistance is achieved when the ordered phase domains in such alloys form a hierarchical mixture of B2 and L2$_1$ subdomains~\cite{Liebscher2013, Sun2013, Liebscher2015, Song2017, Song2017a, Baik2018}.\\

\section{\label{sec:methods}Experimental Methods}

Bulk alloys were produced by vacuum induction casting of 99.9\% purity base elements into water cooled cylindrical copper moulds. The cylindrical cast rods had nominally 25 mm diameter. Sections of rod were removed for further processing and analysis by wire electrical discharge machining (wire EDM). Circular cross section samples with nominally 1.5 mm thickness were removed from a region central to the rod length for preparation of samples to be analysed by X-ray diffraction (XRD) and scanning electron microscopy (SEM). In addition, thin disks were removed at several locations along the rod for wet chemical composition analysis to ensure adherence to the nominal composition and that the elements were homogeneously distributed along the length of the rod. Sections of cast rod with 65 mm length were removed for annealing heat treatment. Annealing was carried out at 900 $^{\circ}$C for 100 h in an Ar atmosphere, followed by furnace cooling.\\

Both as-cast and annealed samples with surface normal parallel to the cast rod cylinder axis were prepared for XRD and SEM analysis by surface grinding using SiC abrasive paper, followed by diamond polishing, and finally by vibratory polishing using a colloidal silica oxide polishing suspension with a mean particle diameter of 50 nm. XRD was carried out with a Bruker D8 Advance instrument in Bragg-Brentano geometry and with Co as the X-ray source. Rietveld refinement of the XRD patterns was performed using the MAUD software package to determine lattice constants and estimate phase volume fractions. Samples were imaged in a Zeiss Sigma field emission SEM using primarily backscattered electron (BSE) imaging contrast. The BSE imaging was carried out with a four quadrant solid-state detector. Energy Dispersive Spectroscopy (EDS) was also performed in the SEM, using an EDAX Octane Elect silicon drift detector. SEM-EDS quantification was carried out using standardless eZAF correction with the EDAX APEX\textsuperscript{\textregistered} software.\\

BSE images were analysed using in-house Python programs which made use of a watershed image segmentation algorithm~\cite{Beucher1993} for determination of average equivalent planar particle radii, $\langle r_s\rangle$, and areal number density, $N$, for regions of the samples with equiaxed precipitates in a matrix phase. Precipitate volume fractions, $f$, and equivalent true radii, $\langle r\rangle$ were calculated making use of stereological corrections according to~\cite{Ardell1985},\\
\begin{equation}
    \langle r\rangle=\frac{4}{\pi}\langle r_s\rangle,\\
\end{equation}
\begin{equation}
    f=\frac{32}{3\pi}\langle r_s\rangle^2N.\\
\end{equation}
Adaptive thresholding was used for determination of area fractions of ordered and disordered phases in regions of material with an interpenetrating maze-like morphology. Phase volume fraction in these regions was assumed equivalent to the measured area fraction. Average domain width in maze-like regions was determined by manual measurement.\\

Samples for scanning transmission electron microscopy (STEM) analysis were primarily prepared using a Thermo Fisher Scientific SCIOS\textsuperscript{\textregistered} DualBeam\textsuperscript{\textregistered} SEM and focused ion beam (FIB) system equipped with Ga$^+$ ions. Cross section specimens were removed from the previously prepared SEM samples by milling trenches using an ion beam accelerating voltage of 30 kV and currents between 50 nA and 300 pA. Final thinning of the TEM lamellas to $\approx$50 nm thickness was carried out using 2 kV and 35 pA ion beam accelerating voltage and current. One crept specimen was prepared by electropolishing using a Struers Tenupol-5 with 22 V potential and 16 $^{\circ}$C in a 10\% perchloric acid-90\% acetic acid electrolyte.\\

High resolution STEM was performed using both probe and image aberration corrected FEI Titan Themis 60-300 instruments (Thermo Fisher Scientific) operated at an acceleration voltage of 300 kV. STEM-EDS was conducted using four synchronised EDS detectors (ChemiSTEM system, Thermo Fisher Scientific). For STEM imaging and EDS, a semi-convergence angle of 23.8 mrad was used. Several high angle annular dark-field (HAADF) detector camera lengths were used in this study; the majority of HAADF images were collected with collection angles of 73 to 200 mrad, while the images in~\cref{fig:d2_crept}(e) and (f) result from collection angles of 25 to 154 mrad and are thereby referred to as low-angle annular dark-field (LAADF). All other LAADF images were acquired with collection angles of the annular dark field detector of 18 to 73 mrad. The beam current was set to approximately 500 pA for STEM-EDS and lowered to 80 pA for image acquisition. Dark-field imaging shown in inset images of~\cref{fig:STEM}(d) were acquired using a JEOL 2100Plus operated at 200 kV acceleration. Non-negative matrix factorization principle component analysis (NMF-PCA) was carried out using the open-source electron microscopy analysis python package, HyperSpy
~\cite{Delapena2020}.\\

Tensile testing was carried out using a Zwick/Roell Z100 universal testing machine. All tests were conducted under displacement control with a constant strain rate of $10^{-4}$ s$^{-1}$, using the laserXtens 2-120 HP/TZ laser extensometer. Rectangular cross-section specimens were machined using wire EDM followed by surface grinding to remove surface oxide. Specimens had dog-bone geometry with 2 mm thickness, 5 mm width in the gauge-length, 30 mm gauge-length, 5 mm radius to grips. Specimens were positioned in form-fit grips. A three zone high temperature furnace was used for specimen heating, with dual thermocouples placed directly on the specimen surface and digital temperature control. Samples were tested in air with a maximum heating rate of 20 K/min.\\

Static creep testing was performed using live image tracking with a video extensometer on a purpose-built testing apparatus, as described in~\cite{Volkl2003, Volkl2008}. Specimens were tested in air using a radiation furnace, with temperature measured using an S-type thermocouple directly in contact with the specimen surface. Rectangular cross-section specimens with 1 mm thickness, 2 mm gauge width and 8 mm gauge length were tested in form-fit grips. Samples were loaded only after the target temperature had been reached and was observed to be stable within $\pm5$ $^{\circ}$C. Both single loading and step loading tests were conducted. For single loading tests, a constant load was hung from the sample and the sample was either crept until failure or interrupted once a constant steady-state creep rate was observed by turning off the furnace and letting the sample cool under load. Step loading tests were conducted by allowing the sample to reach a steady-state creep rate at the initial load and then adding successive load only after the creep rate was observed to reach a new steady-state. Data points from step loading tests were only considered to be valid if the strain rate at the previous load step was not observed to have increased prior to the load addition.\\

\section{Results}

The nominal and measured (wet chemical) compositions of the three cast alloys studied here are listed in~\cref{tbl:comps}, along with an estimate of their densities based on a weighted sum of the nominal pure elemental components. The measured compositions for all elements deviate by no more than 0.4 at.\% from the nominal values, which is acceptably near to refer to the nominal values throughout the rest of the article.\\

\begin{table}[h]
\centering
\caption{\label{tbl:comps}Nominal and measured compositions (in at.\%) and calculated densities of alloys produced by vacuum induction casting. Uncertainties in the measured compositions indicate the experimental measurement error.}
\begin{tabu}{c | c | c c c c c | c c c c c}\toprule
\multirow{2}{*}{Sample ID} & Density & \multicolumn{5}{c|}{Nominal Composition} & \multicolumn{5}{c}{Measured Composition}\\
& [g/cm$^3$] & Al & Cr & Fe & Ni & Ti & Al & Cr & Fe & Ni & Ti\\\hline
D1 & 6.6 & 16 & 25 & 35 & 20 & 4 & 15.9$\pm$0.1 & 24.9$\pm$0.1 & 35.1$\pm$0.2 & 19.9$\pm$0.2 & 4.22$\pm$0.05\\
D2 & 6.4 & 20 & 20 & 35 & 20 & 5 & 19.8$\pm$0.1 & 20.2$\pm$0.1 & 35.0$\pm$0.2 & 20.0$\pm$0.2 & 5.03$\pm$0.05\\
D3 & 6.1 & 24 & 15 & 35 & 20 & 6 & 23.7$\pm$0.1  & 15.1$\pm$0.1 & 35.4$\pm$0.2 & 19.8$\pm$0.2 & 5.94$\pm$0.05\\
\bottomrule
\end{tabu}
\end{table}

\subsection{Phase and Bulk Lattice Parameter Determination}

\begin{figure}[H]
\centering
\captionsetup{skip=4pt}
\includegraphics[width=0.9\textwidth]{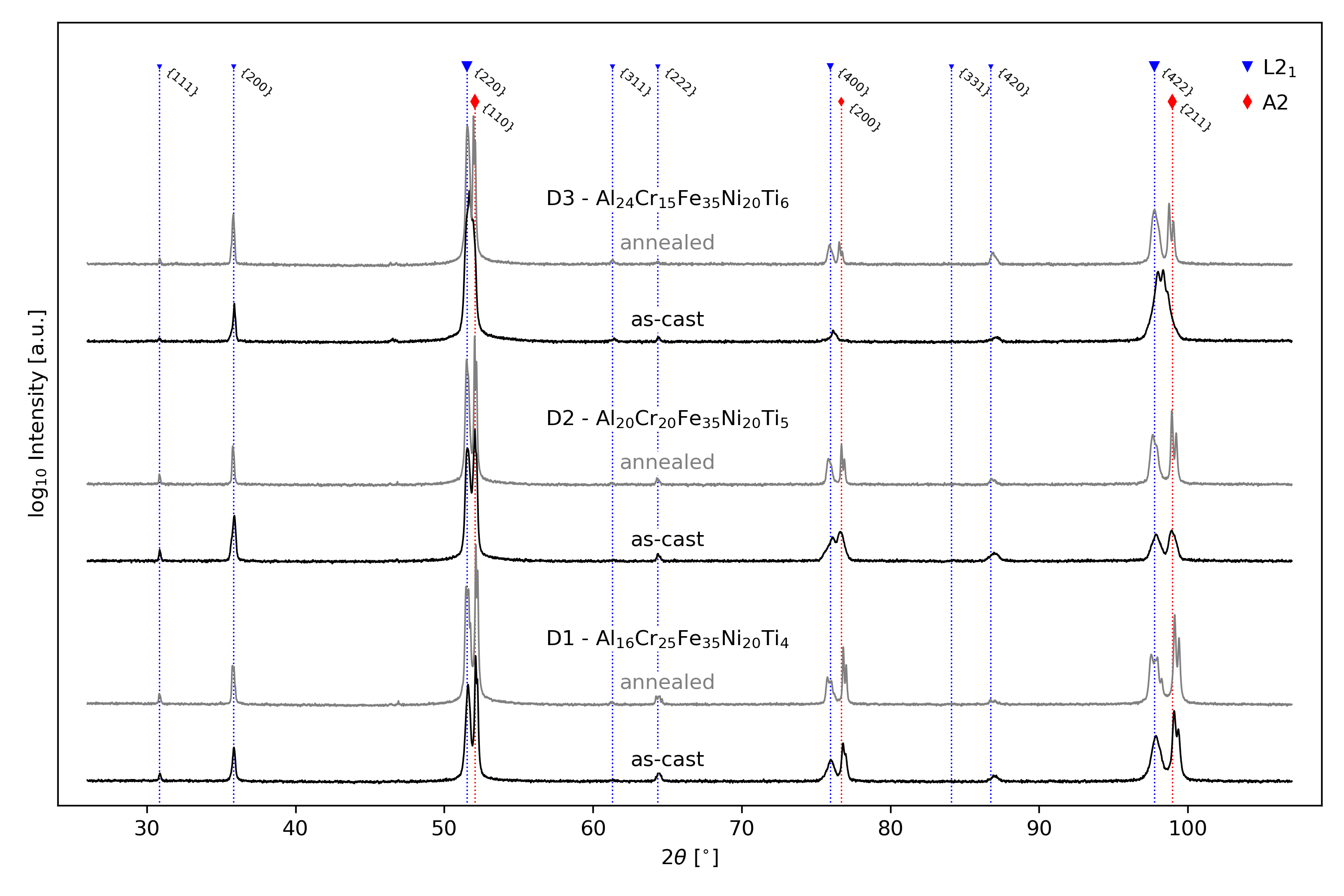}
\caption{\label{fig:xrd}XRD patterns for all three compositions in both the as-cast and 100 h annealed conditions.}
\end{figure}

We conducted XRD to identify the phases that formed in both the as-cast and annealed conditions. The obtained patterns are shown in~\cref{fig:xrd} over the scan range $2\theta \in 25^{\circ} - 105^{\circ}$.  We have conducted a Rietveld refinement of the patterns, and found that they are fitted well by A2 ($\alpha$-Fe) and L2$_1$ ($\beta'$-Ni$_2$TiAl) phases. We note that the B2 ($\beta$-NiAl) phase could also be used in fitting the patterns, however, due to the similarity between B2 and L2$_1$ phases, as well as the difficulty in distinguishing their contributions from XRD analysis alone, we have chosen to fit just the L2$_1$-ordered phase. During the fitting process we refined mainly the lattice parameter to achieve the best fit for all peaks in the scan range $2\theta \in 25^{\circ} - 130^{\circ}$. From the patterns we note that the splitting between individual A2 and L2$_1$ primary peaks (due to different lattice parameters), which is most prominently seen for the A2-$\left\{211\right\}$/L2$_1$-$\left\{422\right\}$ peaks, increases from alloy D1 to alloy D3, especially in the as-cast patterns. For the D1 alloy in the annealed condition we observe a noticeable splitting of the ordered phase peaks, especially for the $\left\{422\right\}$ reflection. This could indicate the presence of multiple ordered phase variants in this alloy, which might arise due to compositional heterogeneities in the sample.\\

Lattice parameters extracted from the refined patterns are listed in~\cref{tbl:xrd}. The domain averaged A2 lattice parameter is found to slightly increase from alloy D1 to D3---that is, with increasing Al and Ti content and decreasing Cr content---for both as-cast and annealed conditions. For all three alloys the A2 lattice parameter does not appear to change significantly upon annealing. Directly comparing the L2$_1$ phase lattice parameter for the three alloys in both the as-cast and annealed state, one observes effectively the same lattice constant for all three alloys, with a slightly lower value for alloy D3. One also observes a slight increase in L2$_1$ lattice parameter after annealing heat treatment\\

Of great importance to the phase morphology in these alloys, and thereby to the high temperature mechanical properties, is the misfit between the ordered and disordered phase lattice parameters. This misfit is calculated as,\\
\begin{equation}
\delta_{L2_1-A2}=\frac{\frac{a_{L2_1}}{2}-a_{A2}}{0.5(\frac{a_{L2_1}}{2}+a_{A2})}.\\
\end{equation}
and values for the three alloys have also been included in~\cref{tbl:xrd}. In both as-cast and annealed conditions, the misfit is seen to decrease from approximately 1\% in alloy D1 to approximately 0.62\% in alloy D3. In other words, the lattice mismatch decreases with increasing Al and Ti and decreasing Cr content. Finally, we observe that the lattice parameter misfit increases upon annealing for all three alloys.\\

\begin{table}[h]
\centering
\caption{\label{tbl:xrd} Domain averaged lattice parameters from refined XRD patterns in both as-cast and annealed conditions, as well as calculated lattice misfit values. Values within parentheses are non-significant.}
\begin{adjustbox}{max width=\textwidth}
\begin{tabu}{l | c c | c c | c c}\toprule
\multirow{2}{*}{ID - Composition} & \multicolumn{2}{c|}{$a_{A2}$ [$\textrm{\AA}$]} & \multicolumn{2}{c|}{$a_{L2_1}/2$ [$\textrm{\AA}$]} & \multicolumn{2}{c}{$\delta_{L2_1-A2}$ [\%]}\\
& as-cast & \textcolor{mygray}{annealed} & as-cast & \textcolor{mygray}{annealed} & as-cast & \textcolor{mygray}{annealed}\\\hline
D1 - \One & 2.87(9) & \textcolor{mygray}{2.87(9)} & 2.90(8) & \textcolor{mygray}{2.91(2)} & 1.00 & \textcolor{mygray}{1.13}\\
D2 - \Two & 2.88(4) & \textcolor{mygray}{2.88(3)} & 2.90(9) & \textcolor{mygray}{2.91(1)} & 0.86 & \textcolor{mygray}{0.97}\\
D3 - \Three & 2.88(8) & \textcolor{mygray}{2.88(7)} & 2.90(6) & \textcolor{mygray}{2.90(9)} & 0.62 & \textcolor{mygray}{0.77}\\
\bottomrule
\end{tabu}
\end{adjustbox}
\end{table}

\subsection{Microstructural Characterisation}

\begin{figure}[h!]
\centering
\captionsetup{skip=4pt}
\includegraphics[width=\textwidth]{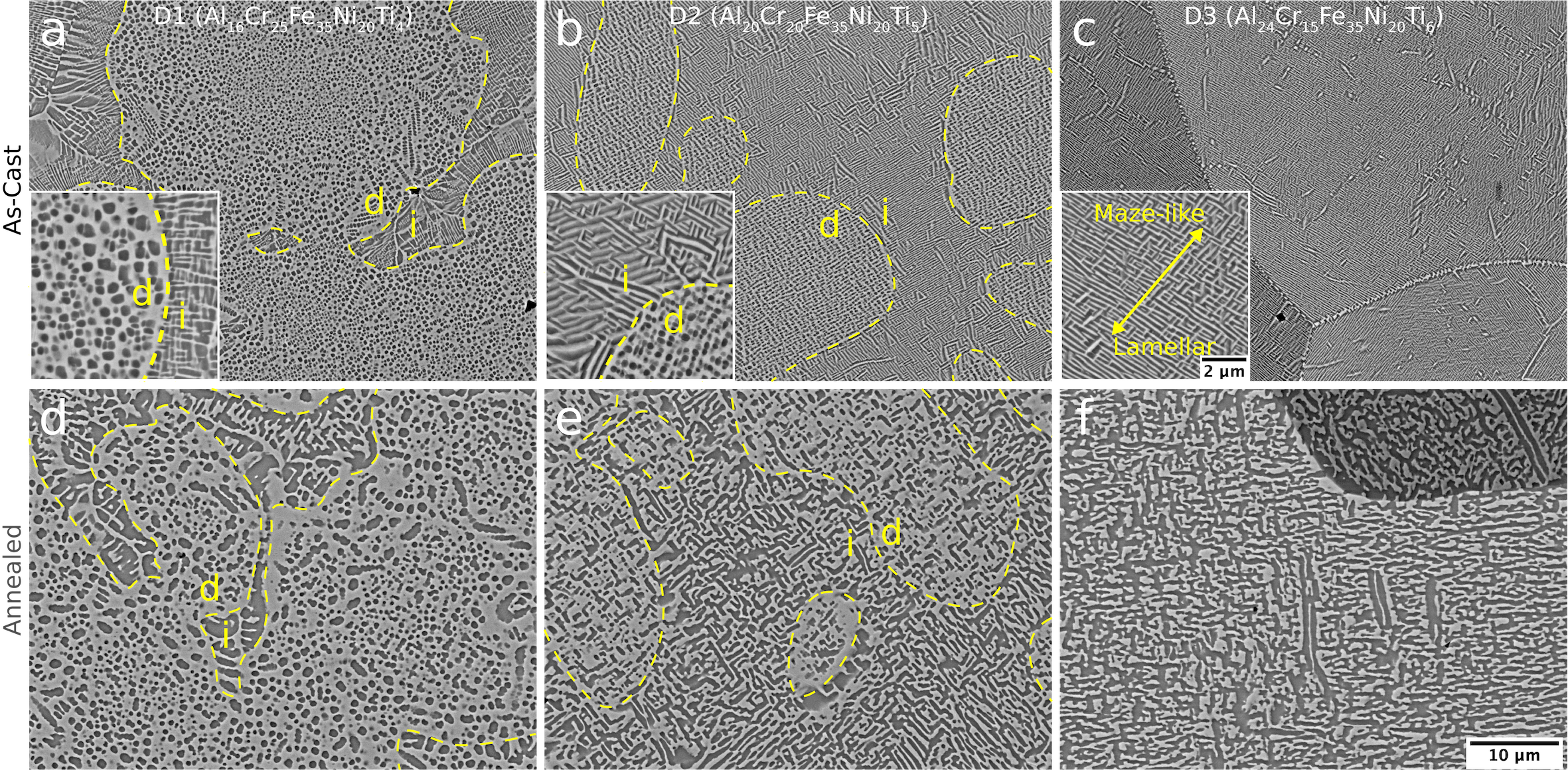}
\caption{\label{fig:SEM}BSE SEM images of alloys D1-D3 in both annealed and as-cast conditions with same scale across all images. Inset in as-cast images shows higher magnification of representative region, with same scale across all insets. In all images, bright domains have been identified as the A2 phase, and dark domains as B2/L2$_1$. Dendritic (d) and interdendritic (i) regions have been identified for alloys D1 and D2. A region with more maze-like appearance and one with more lamellar appearance are indicated in the inset for as-cast alloy D3.}
\end{figure}

To characterise the segregation behavior of alloys D1-D3 in the as-cast and annealed conditions we employed SEM-EDS. \cref{fig:SEM} shows EDS elemental mappings of the five constituent elements for all three alloys in both conditions. One observes clearly defined dendritic segregation structures in all five elemental maps for alloys D1 and D2, with much less noticeable segregation also present in alloy D3. The primary dendrites in both D1 and D2 are enriched in Fe and Cr and depleted in Al, Ni, and Ti, indicating that the primary phase to form was of A2 structure. Inverse segregation behavior is seen in interdendritic regions. The primary dendrites occupy a much larger portion of the sample in alloy D1 than in alloy D2, where the primary dendrites occupy less area than the interdendritic region. Noteworthy are the maps of Ti content, where for alloy D1 the dendritic structures are less defined than for the other elements, and for alloy D2, there is a nearly continuous gradient of Ti content from the dendrite cores to the grain boundaries. Alloy D3 shows the opposite segregation behavior as alloys D1 and D2, with slight dendritic enrichment of Al and Ni, and depletion of Fe and Cr. The exception is Ti, which appears to be depleted in the dendrite cores but which, again, shows a nearly continuous positive gradient towards the grain boundaries.\\

\begin{figure}[h!]
\centering
\captionsetup{skip=4pt}
\includegraphics[width=0.95\textwidth]{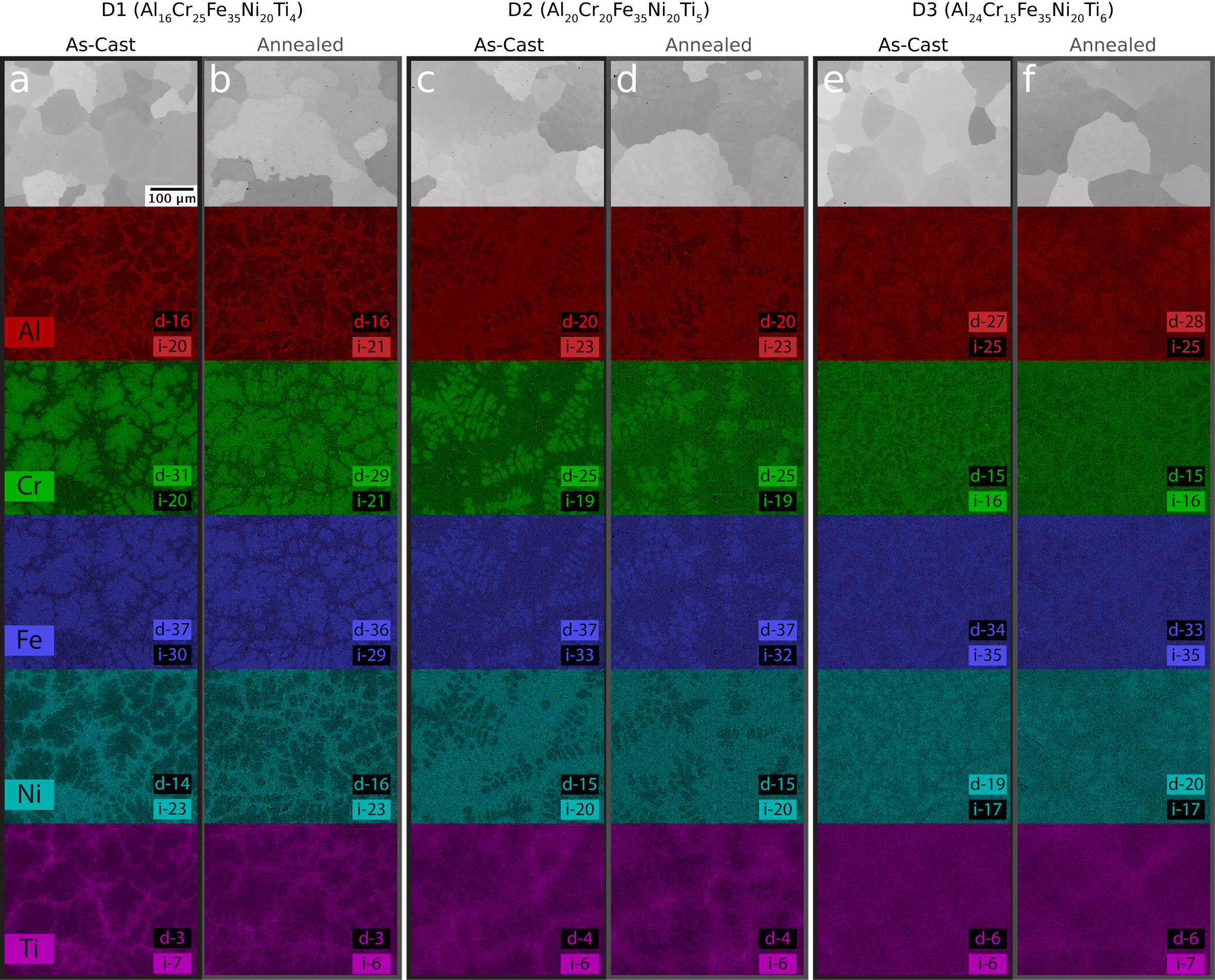}
\caption{\label{fig:sem_eds}Low magnification SEM-EDS mappings of alloys D1-D3 in both annealed and as-cast conditions. Elemental concentrations (in at.\%) are included in each map for both dendritic (d-) and interdendritic (i-) regions. Elemental depletion is indicated by colored text on a black background, and enrichment by black text on a colored background.}
\end{figure}

Quantification of the average elemental concentrations within the dendritic versus interdendritic portions of the alloys is also indicated in~\cref{fig:sem_eds}. From the values in the figure it is clear that the magnitude of segregation decreases for all elements from alloy D1 to alloy D3. As this behavior can have many different causes, it is important to consider the relative degree to which each element segregates. For as-cast alloys D1 and D2, Ti shows the strongest segregation, with 4 at.\% lower Ti content in the D1 dendrites (where the nominal alloy contains 4 at.\% Ti), and 2 at.\% lower D2 dendrites (where the nominal alloy contains 5 at.\% Ti). Ni and Cr show the second strongest segregation (their behavior being nearly perfectly inverse), with D1 dendrites having 11\% greater Cr (nominal 25\%) and 9\% lower Ni (nominal 20\%) than the interdendritic regions, and D2 dendrites having 6\% greater Cr (nominal 20\%) and 5\% lower Ni (nominal 20\%) than the interdendritic regions. Al shows marginally less segregation, with D1 primary dendrites having 4\% lower Al content than interdendritic regions (nominal 16\%), and D2 primary dendrites having 3\% lower Al content than interdendritic regions (nominal 20\%). Finally, Fe shows the lowest degree of dendritic segregation, with D1 dendrites having 7\% greater Fe content and D2 dendrites having 4\% greater Fe content (both with nominally 35\%). For both alloys, this behavior remains the same after annealing, but the degree of segregation decreases slightly for alloy D1. For alloy D3, Ni segregates the most (2\% greater in dendrites, with 20\% nominal), followed by Al (2\% greater in dendrites, 24\% nominal), Cr (1\% lower in dendrites, 15\% nominal), Fe (1\% lower in dendrites, 35\% nominal), and finally Ti (same content in dendritic and interdendritic regions). While annealing appears to also have little effect on the overall segregation trends in alloy D3, we do observe a slight increase in segregation magnitude. Although the quantified elemental concentrations give us a good indication of the trends across the three alloys, we note the prevalence of several systematic errors. Namely, the quantified Al content in all six mappings is between 1 and 2 at.\% higher than expected from the wet chemical composition results, the Ni content is between 1 and 2 at.\% lower than expected, and the Cr and Fe contents are about 1 at.\% higher and lower then expected, respectively.\\

With the systematic errors taken into account, the findings indicate that Cr content is the main driver for segregation in these alloys, with Cr showing poor solubility in Ni rich regions and vice versa. Furthermore, Ti appears to show low solubility in Cr rich regions, but also shows a tendency across all alloys to become increasingly enriched towards the grain boundaries. In alloy D3, this behavior is the least pronounced, reflecting the expected behavior that Ti has a greater solubility in Ni-Al rich (and thereby more ordered phase) regions. The strong dendritic segregation in alloys D1 and D2 is expected to result in very different phase fractions and morphologies in the dendritic and interdendritic regions of these alloys, with high Fe and Cr content in the primary dendrites suggesting that a high volume fraction of A2 phase should be present in these regions. The opposite is true in the interdendritic regions of D1 and D2, where high Al, Ni, and Ti content should lead to a high ordered phase volume fraction. For alloy D3, a high ordered phase volume fraction is expected throughout the structure. Higher resolution characterisation is necessary to investigate the respective phase fractions further.\\

BSE imaging contrast was chosen to characterise the microstructural morphologies of the alloys at higher magnification in the SEM, as this imaging mode allows efficient and clear identification of the A2 and B2/L2$_1$ phase domains due to the different average atomic number of the ordered and disordered phases~\cite{Wolff-Goodrich2021}. In all BSE images, disordered domains appear with bright contrast, having higher average atomic number than ordered domains, which appear dark. BSE SEM images of samples before and after annealing heat treatment are shown in~\cref{fig:SEM}. For alloys D1 and D2, the elemental segregation that was observed in~\cref{fig:sem_eds} is resolved and reveals distinct phase morphology in the dendritic and interdendritic areas.\\

For alloy D1, the primary dendrites are made up of an A2 matrix with equiaxed precipitates ranging from 120 to 700 nm in diameter in the as-cast state and from 0.3 to 2 $\mu$m in the annealed state. As-cast, most precipitates show a cuboidal shape with strong faceting in certain orientations. The ordered precipitates make up $\approx$ 33\% (volume fraction, determined as described in~\cref{sec:methods}, Experimental Methods) of the dendritic regions as-cast and $\approx$ 30\% in the annealed condition. The interdendritic regions contain alternating domains of A2 and B2/L2$_1$ structure which have a lamellar arrangement which is perpendicular to the dendrite-interdendrite interface, and grows coarser towards the grain boundaries. The A2 domain width varies between about 100 and 800 nm in the as-cast state and between about 0.2 and 2 $\mu$m after annealing, with ordered phase domains being larger than disordered domains in both conditions. In some especially coarse ordered phase domains in the as-cast state, very fine particles and needles of the disordered phase can be observed, and likely formed by a secondary decomposition reaction upon cooling. In the interdendritic regions, the L2$_1$ precipitates make up  $\approx$ 59\% of structure in the as-cast state and $\approx$ 57\% in the annealed state.\\

For alloy D2, the primary dendrites are also made up of an A2 matrix with equiaxed cuboidal precipitates, however, these are somewhat finer than in the D1 alloy, ranging from 100 to 500 nm in diameter in the as-cast state and from 0.2 to 2 $\mu$m in the annealed state. Similar to D1, the precipitates in the dendritic region show faceted interfaces with the disordered matrix. After annealing, the precipitates are no longer mostly equiaxed, showing irregular shapes in some grains, possibly due to particle coalescence. The ordered precipitates make up $\approx$ 34\% of the dendritic regions in the as-cast condition and $\approx$ 30\% in the annealed condition. The interdendritic regions, which make up the majority of the alloy volume, contain a fine maze-like arrangement of A2 and B2/L2$_1$ phases. The A2 domain width in these regions ranges from about 60 and 380 nm in the as-cast state and between about 0.12 and 1.2 $\mu$m in the annealed state, with the A2 domains being slightly narrower than the ordered phase domains. Within the interdendritic regions, the ordered phase comprises $\approx$ 55\% of the structure in both the as-cast and annealed states.\\

For alloy D3, it is very difficult to distinguish between dendritic and interdendritic regions from SEM images alone. There is a tendency for regions identified as dendritic to have a more maze-like appearance, and for regions identified as interdendritic to have a more lamellar appearance, however, this is difficult to distinguish, especially at low magnifications. Furthermore, there is no strict boundary between regions of maze-like and lamellar appearance, as for alloys D1 and D2, but rather a gradient between the regions, as can be seen in the inset in~\cref{fig:SEM}(c). The A2 domain width in this alloy ranges between 50 and 250 nm in the as-cast state, and between 0.1 to 0.9 $\mu$m in the annealed state. The ordered phase makes up $\approx$ 58\% of the structure both as-cast and after annealing.\\

From our observations of the phase morphologies we see that all three alloys coarsen at similar rates. Furthermore, coarsening in the primary dendrites of alloys D1 and D2 appears to occur slower than the coarsening in the maze-like/lamellar regions of alloys D2 and D3. While the interdendritic regions of alloy D1 appear related to those in alloy D2 and the structure throughout alloy D3, they have a slightly coarser appearance which is reminiscent of eutectic phase formation, while the structures in D2 and D3 are more likely to have formed by a solid-state decomposition mechanism.\\

\subsection{As-cast Phase Chemistry and Interface Structure}

\begin{figure}[h!]
\centering
\captionsetup{skip=4pt}
\includegraphics[width=0.95\textwidth]{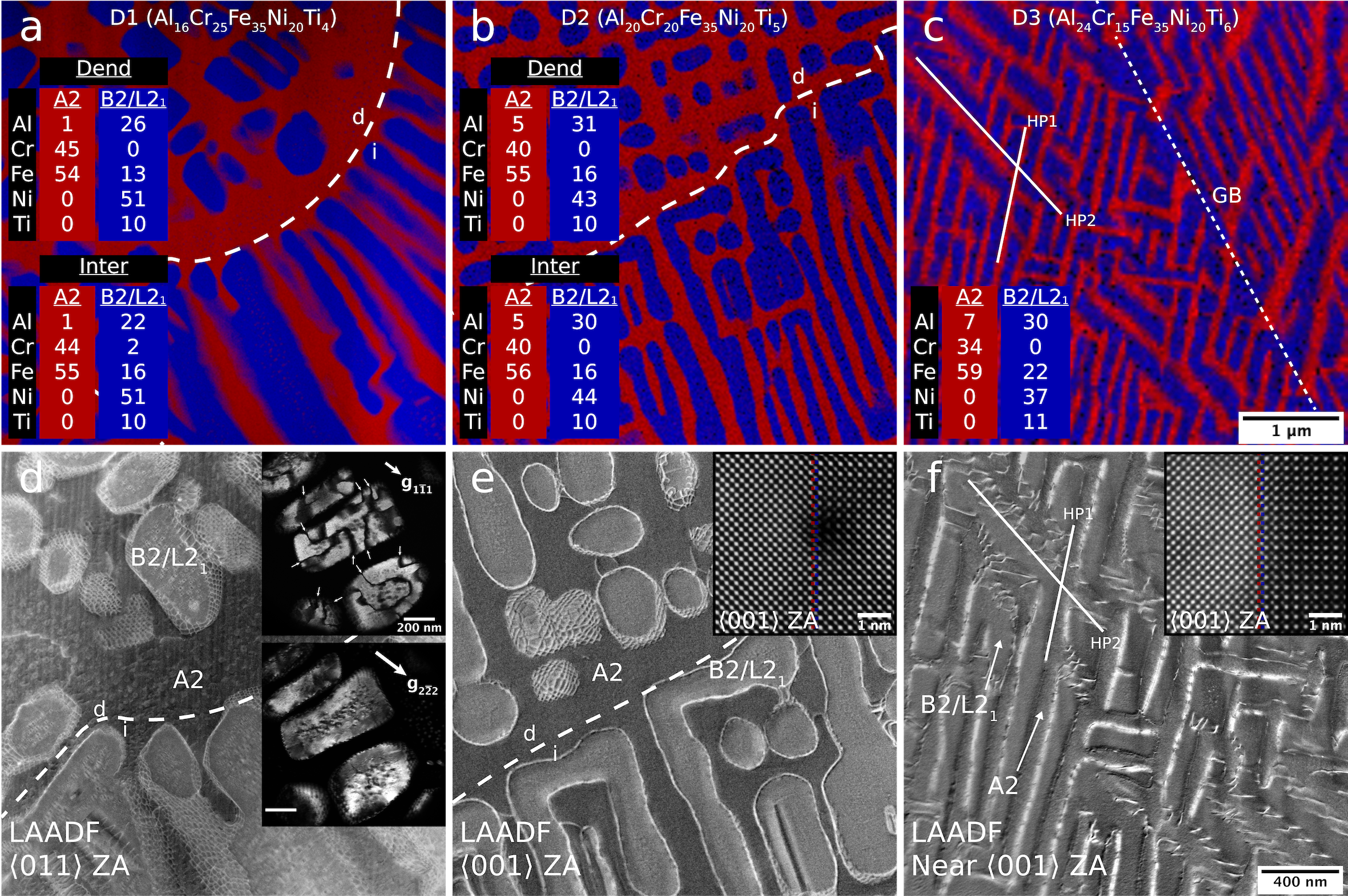}
\caption{\label{fig:STEM}First row images show loadings from NMF-PCA decomposition of STEM-EDS maps of D1-D3 in the as-cast condition. Quantified phase compositions (in at.\%) are indicated directly in the images for ordered (B2/L2$_1$) and disordered (A2) phases. Scale for all three maps is the same. Second row images show phase structure and interfacial dislocation networks, imaged using a low angle annular dark-field (LAADF) detector which shows both atomic number and diffraction contrast. Scale for all three images is the same. Inset dark field images in (d) were taken  from the same area with $g_{1\overline{1}1_{L2_1}}$ and $g_{2\overline{2}2_{L2_1}}$ diffraction vectors. Locations of superlattice APBs are indicated with white arrows for $g_{1\overline{1}1_{L2_1}}$. Inset HAADF images in (e) and (f) have been Fourier filtered to remove high frequency noise, and each show a single interface between ordered and disordered phase domains in alloys D2 and D3.}
\end{figure}

Higher resolution observations of the phase chemistry and interface structure between A2 and B2/L2$_1$ phases have been captured using STEM-EDS mapping and low angle annular dark field (LAADF) diffraction contrast imaging, as shown in~\cref{fig:STEM}. Principle component analysis (PCA) based on non-negative matrix factorization (NMF) has been used to decompose the spectral images into two unique components, which coincide with the ordered and disordered phases (labeled as A2 and L2$_1$). The chemical composition of these components is displayed in~\cref{fig:STEM} for dendritic and interdendritic regions in alloys D1 and D2, and for the entire structure in alloy D3. We have highlighted the boundary between dendritic and interdendritic regions in (a) and (b), as well as the location of a grain boundary (GB) in (c). Note that due to the approximating nature of taking a finite number of components from the decomposition of the spectra, low elemental contributions, which appear as noise in the elemental maps, do not appear in the final compositions. The quantified compositions are therefore mainly useful in comparing trends between the three alloys and in understanding the morphologies which form in different regions of the samples.\\

A clear trend of increasing Al content in the disordered phase is observed, with the alloy D1 A2 phase having about 1 at.\% Al in dendritic and interdendritic regions, and the alloy D3 A2 phase having about 7 at.\% Al. The Cr content in the A2 phase also follows a fairly linear trend, decreasing by about 10 at.\% from alloy D1 to alloy D3, with alloy D2 having an intermediate amount. The Fe content in the A2 phase shows a more complicated relationship, with the alloy D1 and D2 A2 phases showing approximately equal concentrations in the respective dendritic and interdendritic regions, while the alloy D3 A2 phase has more than 3 at.\% greater Fe content than that in alloys D1 and D2. The Al content in ordered phase domains is seen to increase from alloy D1 to alloy D2, however, the Al content in the alloy D3 ordered phase domains is slightly lower than for alloy D2. For all three alloys, the ordered phase domains have non-stoichiometric Al + Ti content, which suggests that there is some degree of Fe and Ni anti-site defects on the Al and Ti sublattice(s)~\cite{Cohron1998, Huffman1967, Wertheim1967, Taylor1972}. In conjunction with this observation we must note, however, that for all elemental maps, a low total Al + Ti content was observed for all B2/L2$_1$ phases, possibly due to a combination of residual matrix phase above or below the precipitates and channeling effects in regions which were aligned near zone-axis orientation~\cite{Kothleitner2014}. Another important trend worth pointing out is the fraction of Fe on the (Fe, Ni) sublattice of the ordered phase. Alloy D1 is seen to have the lowest Fe occupation at about 21\% and 24\% of the (Fe, Ni) sublattice occupation in dendritic and interdendritic regions, respectively. Alloy D2 has a greater fraction, at around 26\% in both dendritic and interdendritic regions. Alloy D3 has the highest with around 38\% Fe on the (Fe, Ni) sublattice. This variation in Fe:Ni occupation occurs despite the fact that the Fe and Ni contents in all three alloys is the same. This can be viewed as arising from the mutual solubility of Fe in the B2 and L2$_1$ phases over a wide temperature and composition range~\cite{Stein2007}. As a simple explanation, the Fe/Ni ratio in the ordered phase increases from alloy D1 to D3 because the Ni content alone is not high enough to accommodate the increasing total volume fraction of ordered phase that is encouraged to form in the alloys as the combined Al and Ti content is increased, thereby necessitating increased Fe occupation in the ordered phase domains.\\ 

The LAADF image in~\cref{fig:STEM}(d) shows dense networks of interfacial dislocations between ordered and disordered phase domains in the D1 sample, reflecting the relatively high lattice parameter mismatch between the ordered and disordered phases, as indicated in~\cref{tbl:xrd}. The LAADF image of alloy D1 also shows contrast variations within the disordered and ordered phases, as well as several dislocations in the bulk of the disordered phase. Higher resolution observations of the spot-like contrast within both phases indicates that secondary phase decomposition has occurred to some extent, with nanometer-scale spherical ordered precipitates forming in the disordered phase, and disordered clusters forming in the ordered phase. The boundaries of the fine disordered clusters in the ordered phase have a bright contrast, indicating either coherency strains or the presence of dislocation loops. In the center-left of the image, one can observe a long dislocation bowing through the disordered phase, with additional bowing points occurring at fine ordered phase precipitates along the dislocation line. The superlattice DF images shown in the inset of~\cref{fig:STEM}(d) show contrast arising exclusively from the L2$_{1}$ ordered phase in the dendritic portion of the D1 alloy. In the top inset image, acquired with the L2$_1$ ($1\overline{1}1$) superlattice reflection, one observes the presence of networks of curved antiphase boundaries (APBs) in the ordered phase domains. These may have been formed as a result of coalescence of individual precipitates, by thermal stresses, or due to rapid continuous ordering from B2 to L2$_1$, as has been suggested in several works where similar domain structures were observed~\cite{Boettinger1988, Bendersky1988, Boettinger1991, Liebscher2013}. In the bottom inset image, taken with a superlattice reflection corresponding to both B2 ($1\overline{1}1$) / L2$_1$ ($2\overline{2}2$), a more or less homogeneous contrast is observed and one can no longer locate the APBs, indicating that the primary precipitates are almost entirely L2$_1$ ordered. One does however observe nanometer scale regions with dark contrast in the DF images, suggesting that secondary A2 precipitates have formed in the L2$_1$ particles.\\

The LAADF image of alloy D2 in the as-cast state,~\cref{fig:STEM}(e), also shows both dendritic and interdendritic regions. The interdendritic region at the bottom half of the image does not show any dislocation networks, however, the interfacial dislocations still occur at the phase boundaries, as indicated by the bright lines separating the ordered and disordered phases. The interfacial dislocation networks are not observed to the same extent in~\cref{fig:STEM}(e) as in (d) due to the different zone axis orientation under which the images were acquired; in (e), the cube-on-cube oriented phase boundaries are primarily viewed edge-on, whereas in (d) they are mostly inclined. One does observe, however, that the dendritic region in the top half of (e) contains several dislocation networks where interfaces between ordered and disordered phases occur within the foil thickness. From these networks, we can comment that the density of interfacial dislocations between ordered and disordered phases in alloy D2 is similar to that observed in alloy D1, reflecting the similar degree of lattice misfit in the two alloys. The atomic resolution HAADF image shown as an inset in~\cref{fig:STEM}(e) shows a mostly coherent interface between ordered and disordered domains in the interdendritic region of the sample, however, we do observe the presence of a single interface dislocation. One also observes several dislocations in the disordered phase, which bow between ordered phase domains. Interestingly, there is little to no nanometer-scale secondary precipitation observable in as-cast alloy D2.\\

Alloy D3 once again shows the finest arrangement of phases, with some domains being separated by only tens of nanometers. The strain contrast of the D3 LAADF image shows a much lower density of interfacial dislocations than for either alloy D1 or D2, again corresponding well with the XRD analysis. This is reflected by the atomically resolved HAADF inset image in~\cref{fig:STEM}(e), where the interface between an ordered and disordered domain is seen to be fully straight and coherent. One observes that the phase arrangement favors two preferred habit planes. One of these is the expected $\left\{001\right\}$ habit plane, designated HP1, which results in interfaces such as that seen in the inset HAADF image. The habit plane which deviates from $\left\{001\right\}$, designated HP2, is inclined relative to the $\langle001\rangle$ zone axis and results in bands decorated with interfacial dislocations. The orientation of these two habit planes has been highlighted in the phase map shown in (c) as well as the LAADF image shown in (f). Similar to alloy D2, alloy D3 shows little to no secondary precipitation.\\


\subsection{Annealed D2 Phase Chemistry and Interface Structure}

\begin{figure}[h!]
\centering
\captionsetup{skip=4pt}
\includegraphics[width=0.32\textwidth]{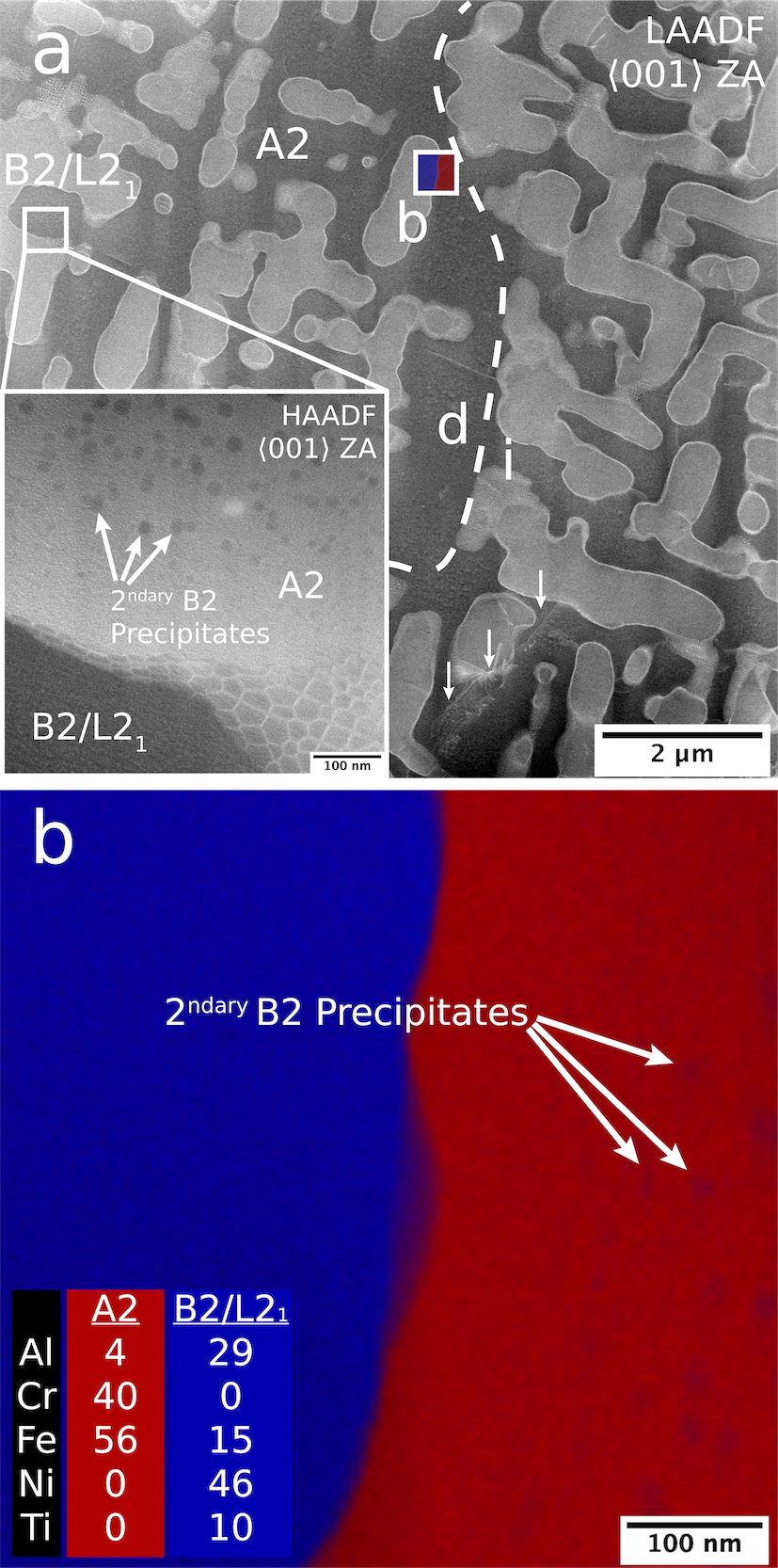}
\caption{\label{fig:d2_ann_STEM}(a) LAADF image of annealed alloy D2 showing both dendritic and interdendritic regions. White arrows indicate region of high dislocation density. HAADF inset in bottom left shows magnified image of phase structure within center of dendritic region. (b) NMF-PCA phase mapping from single interface near the boundary between dendritic and interdendritic regions, indicated in (a).}
\end{figure}

We have investigated the nanoscale structure and chemistry of annealed alloy D2 only, as this alloy consists of regions which are representative of both alloy D1 and alloy D3. An overview LAADF-STEM image is shown in ~\cref{fig:d2_ann_STEM} (a), as well as an inset HAADF-STEM image showing a higher magnification of the structure within individual A2 and B2/L2$_1$ domains. NMF-PCA from EDS mapping of a small portion of the overview image is indicated in (b). We see that the composition in the portion of the annealed sample that we measured is effectively the same as in the as-cast sample. We note that secondary B2 precipitates are discernible in the A2 domain in the EDS mapping, but given their small size, these are largely obscured by the A2 phase above and below. In the HAADF inset we see these secondary precipitates clearly and have determined their diameters to range between about 10 and 25 nm. It is likely that these secondary precipitates formed during cooling from the annealing heat treatment. In the interdendritic region seen in the bottom right of the LAADF overview image, we note a higher number of dislocations in both ordered and disordered phases, with one region (indicated in the image) showing a particularly high density of dislocations in the disordered phase. Closer inspection of this region reveals clear bowing of dislocations around secondary precipitates in the disordered phase. It is not clear why this region has a higher density of dislocations than we otherwise see in the untested as-cast or annealed structures, but may be related to stresses induced during the sample preparation.\\

\subsection{Elevated Temperature Tensile Testing}

\begin{figure}[h!]
\centering
\captionsetup{skip=4pt}
\includegraphics[width=0.95\textwidth]{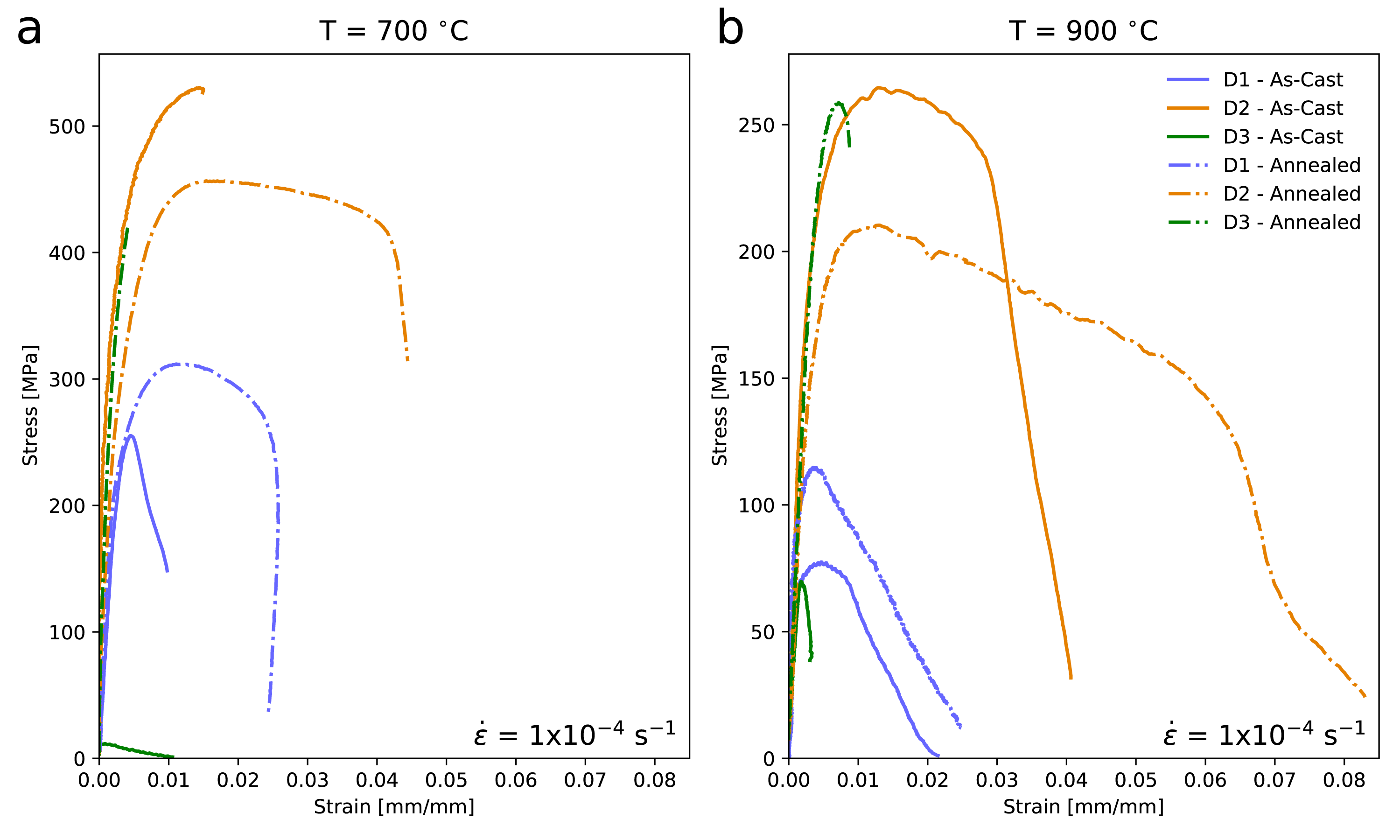}
\caption{\label{fig:tension} Representative tensile engineering stress-strain curves of alloys in both annealed and as-cast conditions tested at 700 $^{\circ}$C (a) and 900 $^{\circ}$C (b). Note that a different stress-axis scaling is necessary for the two test temperatures.}
\end{figure}

Engineering stress-strain curves for all three alloys in both the as-cast and the annealed conditions can be seen in~\cref{fig:tension} for tests conducted at 700 $^{\circ}$C and 900 $^{\circ}$C. The 0.2\% offset yield strength, ultimate tensile strength, and total plastic strain at onset of fracture have been extracted from the stress-strain curves and are indicated in~\cref{tbl:tension}. For tests conducted at 700 $^{\circ}$C, both as-cast alloys D1 and D3 were observed to fracture before yielding over several test attempts, with alloy D3 being particularly prone to brittle fracture at a stress well below the expected yield point of the material. Alloy D2, by contrast, showed yielding behavior in the as-cast state, but fractured before measurable plastic deformation could take place. In the annealed condition, alloy D1 was able to reach yielding and showed a small degree of plastic deformation before failure. Alloy D2 showed nearly the same behavior, but had a significantly higher yield point and slightly greater strain-to-failure. Finally, for alloy D3, it is unclear whether the material was able to reach yielding before failure in the annealed condition, but likely would have had the highest yield point of all the alloys if it had reached yielding.\\

For tests conducted at 900 $^{\circ}$C, the curves appear very similar to those for the 700 $^{\circ}$C tests, but generally reached lower stresses and slightly higher strain-to-failure. Once again, the as-cast alloy D3 failed well before yielding, indicating that even at 900 $^{\circ}$C, this alloy is still extremely brittle. For as-cast alloy D1, it is also unclear whether the material was able to reach yielding before unstable crack growth began, despite conducting multiple tests on the material. In the annealed condition, alloys D1 and D3 both likely reached yielding, but proceeded to fail catastrophically shortly after. The annealed alloy D3 showed the highest yield strength of all three alloys also at this temperature. Alloy D2 showed the highest strain to failure in both as-cast and annealed conditions, and had a yield stress near that of annealed alloy D3.\\

\begin{table}[h]
\centering
\caption{\label{tbl:tension}Results of bulk high temperature tensile tests. Results from samples which failed prior to yielding have been marked with an asterix (*).}
\begin{adjustbox}{max width=\textwidth}
\begin{tabu}{l | c | c c | c c | c c}\toprule
\multirow{2}{*}{ID - Composition} & Test Temp & \multicolumn{2}{c|}{$\sigma_{0.2}$ [MPa]} & \multicolumn{2}{c|}{$\sigma_{uts}$ [MPa]} & \multicolumn{2}{c}{$\epsilon_{plastic}$ [\%]}\\
& [$^{\circ}$C] & as-cast & \textcolor{mygray}{annealed} & as-cast & \textcolor{mygray}{annealed} & as-cast & \textcolor{mygray}{annealed}\\\hline
D1 - \One & 700 & 255* & \textcolor{mygray}{256} & 255* & \textcolor{mygray}{312} & 0.7* & \textcolor{mygray}{2.2}\\
D2 - \Two & 700 & 405 & \textcolor{mygray}{355} & 531 & \textcolor{mygray}{457} & 1.5 & \textcolor{mygray}{4.0}\\
D3 - \Three & 700 & 12* & \textcolor{mygray}{422*} & 12* & \textcolor{mygray}{422*} & 0.0* & \textcolor{mygray}{0.0*}\\\midrule
D1 - \One & 900 & 74* & \textcolor{mygray}{113} & 78* & \textcolor{mygray}{115*} & 0.8* & \textcolor{mygray}{0.5}\\
D2 - \Two & 900 & 223 & \textcolor{mygray}{183} & 265 & \textcolor{mygray}{211} & 2.7 & \textcolor{mygray}{6.1}\\
D3 - \Three & 900 & 70* & \textcolor{mygray}{248} & 70* & \textcolor{mygray}{259} & 0.0* & \textcolor{mygray}{0.8}\\
\bottomrule
\end{tabu}
\end{adjustbox}
\end{table}

\subsection{Fracture Surface Characterisation}

\begin{figure}[h!]
\centering
\captionsetup{skip=4pt}
\includegraphics[width=0.85\textwidth]{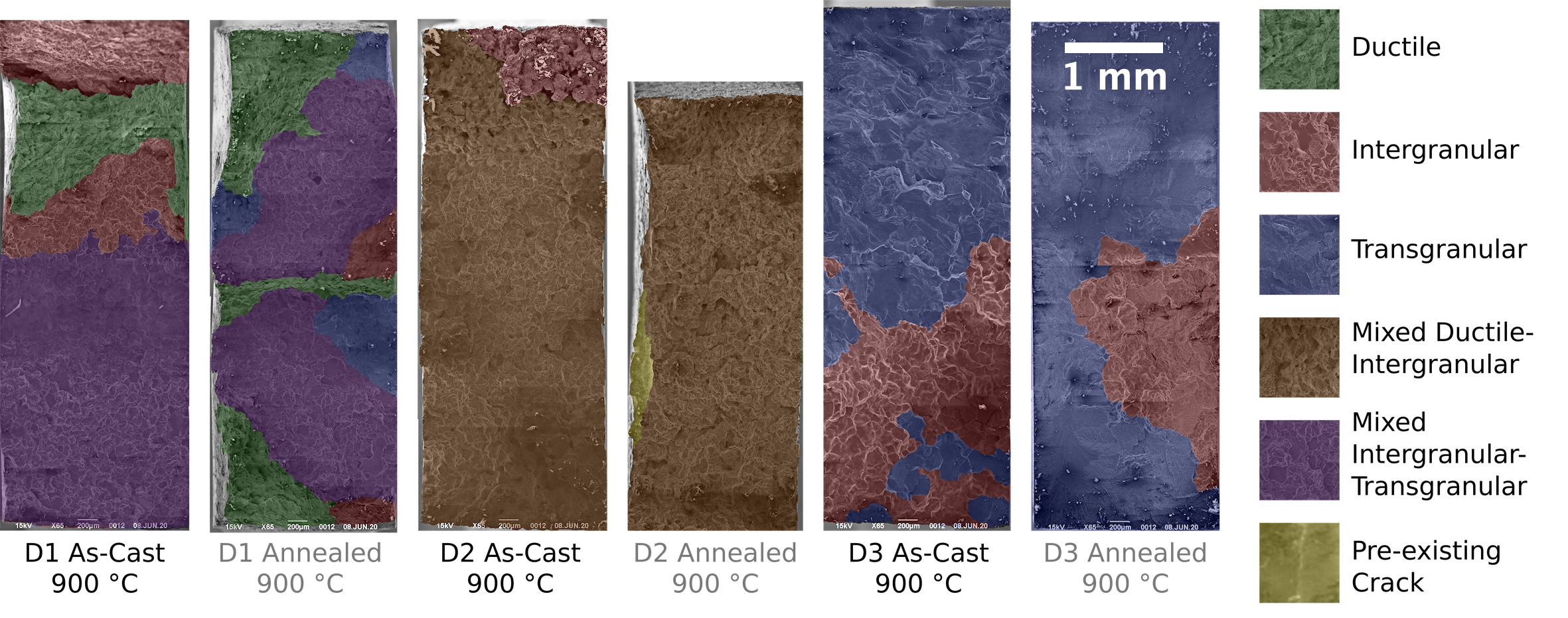}
\caption{\label{fig:fracsurfs}Fracture surfaces of tensile specimens tested at 900 $^{\circ}$C. Colors indicate fracture mode.}
\end{figure}

We conducted post-test fracture surface observations on the samples tested at 900 $^{\circ}$C whose stress-strain curves are shown in~\cref{fig:tension} in order to better understand the failure mode in these alloys. Secondary electron SEM images of the entire fracture surfaces are shown in~\cref{fig:fracsurfs}, with regions of identified fracture mode indicated. Significantly, all three alloys showed some degree of intergranular fracture in both as-cast and annealed conditions. Alloy D1 showed the most complex fracture surfaces, being comprised of regions of pure intergranular type fracture, regions of ductile type fracture, and regions of mixed intergranular-transgranular cleavage type fracture. The as-cast D1 sample tested at 900$^{\circ}$C contains one steep shear wall where the sample contained either very weak grain boundaries or a pre-existing crack prior to testing. This could possibly be the reason that the as-cast sample D1 showed a lower yield strength than the annealed case. The annealed D1 fracture surface shows three regions of ductile failure, but the majority of the fracture surface shows mixed-mode intergranular-transgranular type fracture. Alloy D2, in the as-cast state shows one small region where the sample failed in a purely intergranular manner, but the majority of the fracture surface shows mixed-mode ductile-intergranular fracture. The annealed D2 fracture surface also shows almost entirely ductile-intergranular fracture, and the smaller cross-sectional area of this sample relative to the other samples gives an indication that this sample deformed to the greatest extent before failure. Alloy D3 in both as-cast and annealed conditions shows regions of pure intergranular fracture and regions of purely transgranular cleavage-type fracture. These fracture surfaces are the flattest, reflecting the negligible plastic deformation accommodated by this alloy.\\

\subsection{Creep Tests}



\begin{figure}[h!]
\centering
\captionsetup{skip=4pt}
\includegraphics[width=0.95\textwidth]{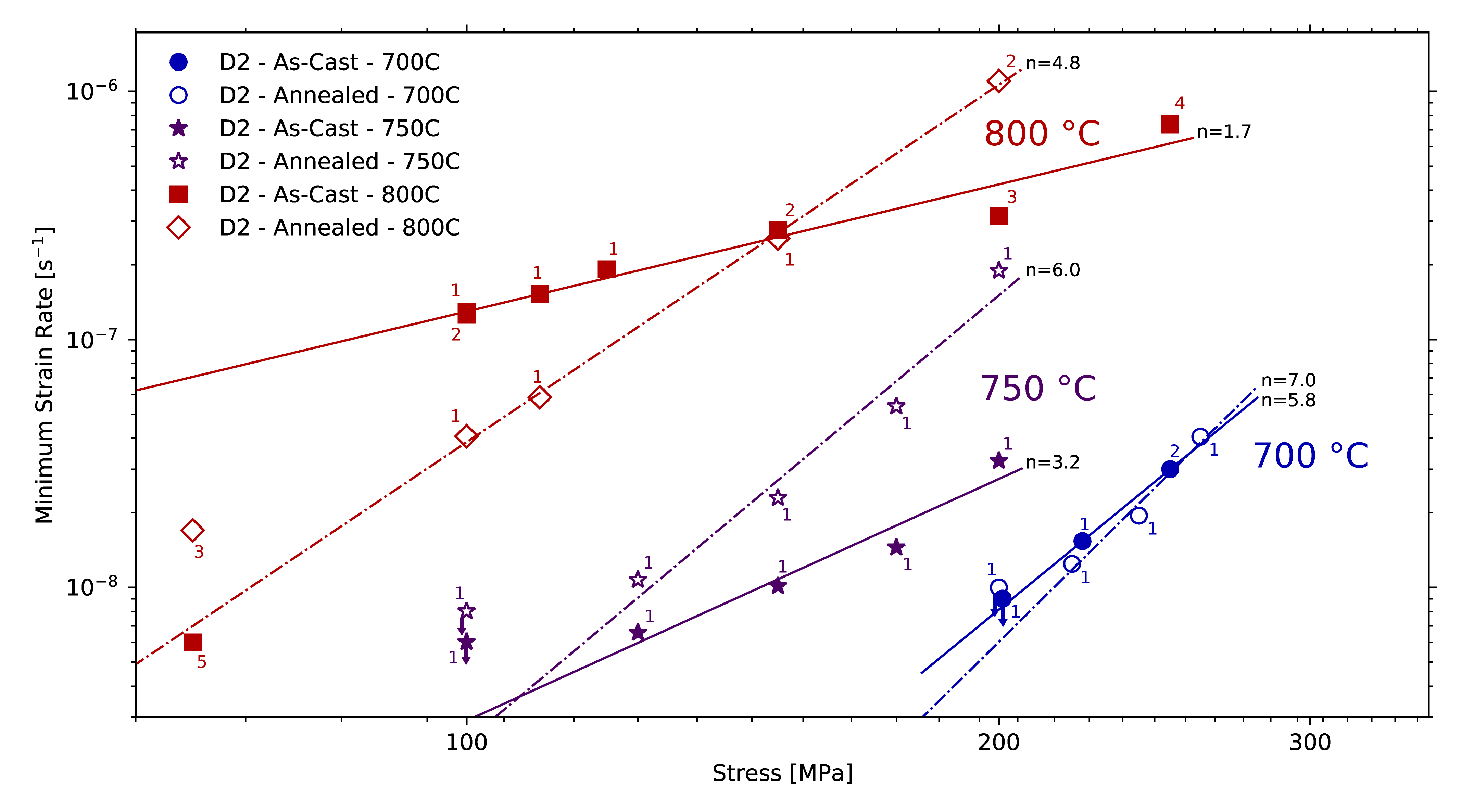}
\caption{\label{fig:d2creep}Steady-state creep rate plot for alloy D2 tested at 700, 750, and 800 $^{\circ}$C. Power-law exponents for each temperature and condition are indicated.}
\end{figure}

Due to the superior combination of moderate strain-to-failure and high tensile yield strength, we chose to conduct creep testing on just alloy D2. Using data from both stepwise-increasing loading tests and normal static loading tests, we extracted the steady-state creep rate of as-cast and annealed alloy D2 under static stresses between 70 and 260 MPa at 700, 750, and 800 $^{\circ}$C. This data is shown on a double logarithmic scale in~\cref{fig:d2creep}. For a given temperature, data points which were taken from a stepwise-loading test all have the same identifying number. As an example, the data points for the as-cast condition, tested at 800 $^{\circ}$C, come from one sample tested under 100, 110, and 120 MPa loading, a second sample tested at 100 and 150 MPa, a third sample tested at 200 MPa, a fourth sample, tested at 250 MPa, and a fifth sample, tested at 70 MPa. Data points from stepwise-loading tests were only included if the strain rate at the previous load step was not observed to increase before the load was increased, and if the strain rate was observed to reach a constant value before the next load increase or sample failure. Assuming power-law creep behavior of the form,\\
\begin{equation} 
\dot{\epsilon}=C\sigma^n,
\end{equation}
we extracted creep exponents for the six sets of data and show these directly in the figure. Note that the logarithm of the minimum creep rate does not vary linearly with the logarithm of stress for all loading conditions. For the 700 $^{\circ}$C annealed sample and the 750 $^{\circ}$C as-cast and annealed samples, we observe that the trend of minimum strain rate follows a concave path with increasing applied stress. Upon re-examining the creep curves, we note that the strain rate at the lowest stresses during these tests was likely lower than what was measured, as we were near the measurement accuracy limit of the testing setup. For this reason, we have not included these lowest stress data points in our linear regression to extract creep exponents. Similarly, the 800 $^{\circ}$C as-cast material which was tested at 70 MPa shows a strong deviation from the behavior at higher stresses, having a strain rate which falls well below the extrapolated power law fit. This behavior suggests that there may be a threshold stress for this material, below which no measurable deformation occurs. The annealed material tested at 70 MPa and 800 $^{\circ}$C shows deviation from the extrapolated fit in the opposite direction. Aside from the deviations just mentioned, we observe that the creep exponent generally decreases with temperature for both as-cast and annealed conditions. Interestingly, we also observe that the creep exponent strongly differs between the as-cast and annealed condition, being lower in the as-cast condition at all temperatures. The very low creep exponent of 1.7 in as-cast D2 at 800 $^{\circ}$C above 100 MPa is suggestive of a diffusion based creep mechanism at this temperature and stress level, however, observations of the as-crept microstructure are needed to confirm this.\\

From the tests conducted at 700, 750, and 800 $^{\circ}$C at 200 MPa loading, we have extracted values of the creep activation energy, $Q$, by taking the slope of the best fit line through an Arrhenius plot of $\ln{\dot{\epsilon}}$ versus $1/T$ for as-cast and annealed strain rate data. For 200 MPa loading, the creep activation energy was found to be 307 kJ/mol for the as-cast material and 410 kJ/mol for the annealed material. The value of 307 kJ/mol coincides exactly with the value of activation energy for $^{63}$Ni diffusion in stoichiometric NiAl~\cite{Hancock1971, Polvani1976}. We cannot comment on whether the activation energy is also in agreement with that for diffusion in L2$_1$-Ni$_2$TiAl, due to the lack of diffusion data for this phase in the literature, but we assume here that the activation energy for Ni diffusion is similar for B2 and L2$_1$ phases. The activation energy for self-diffusion in $\alpha$-Fe is known to exhibit a discontinuity at the Curie temperature (770 $^{\circ}$C in pure $\alpha$-Fe), i.e. the value in the ferromagnetic state (291 kJ/mol) is different than in the paramagnetic state (239 kJ/mol)~\cite{Kuvcera1982}. However, if data from above and below the Curie temperature are evaluated as a 
``linear Arrhenius set", a self-diffusion activation energy of 307 kJ/mol has been found for polycrystalline $\alpha$-Fe in the temperature range 700 to 900 $^{\circ}$C~\cite{Kuvcera1982}. As this temperature range includes the temperatures of the creep experiments used to extract the creep activation energies, we can consider this Fe self-diffusion activation energy to be representative for the A2 phase in our alloys. The close agreement between the activation energy for self-diffusion in $\alpha-Fe$ and Ni diffusion in NiAl corresponds well with our observation that the steady-state creep rate in as-cast D2 is controlled by bulk diffusion. The significantly higher value of creep activation energy in annealed D2 suggests that the controlling creep mechanism is not diffusion-based. We note, however, that the real activation energy values under this stress level may be lower than determined here, given our previous acknowledgement that the strain rates in the as-cast and annealed condition at 700 $^{\circ}$C, 200 MPa loading are likely to be lower than reported.\\

\subsection{Crept Microstructure Analysis}

\begin{figure}[h!]
\centering
\captionsetup{skip=4pt}
\includegraphics[width=0.95\textwidth]{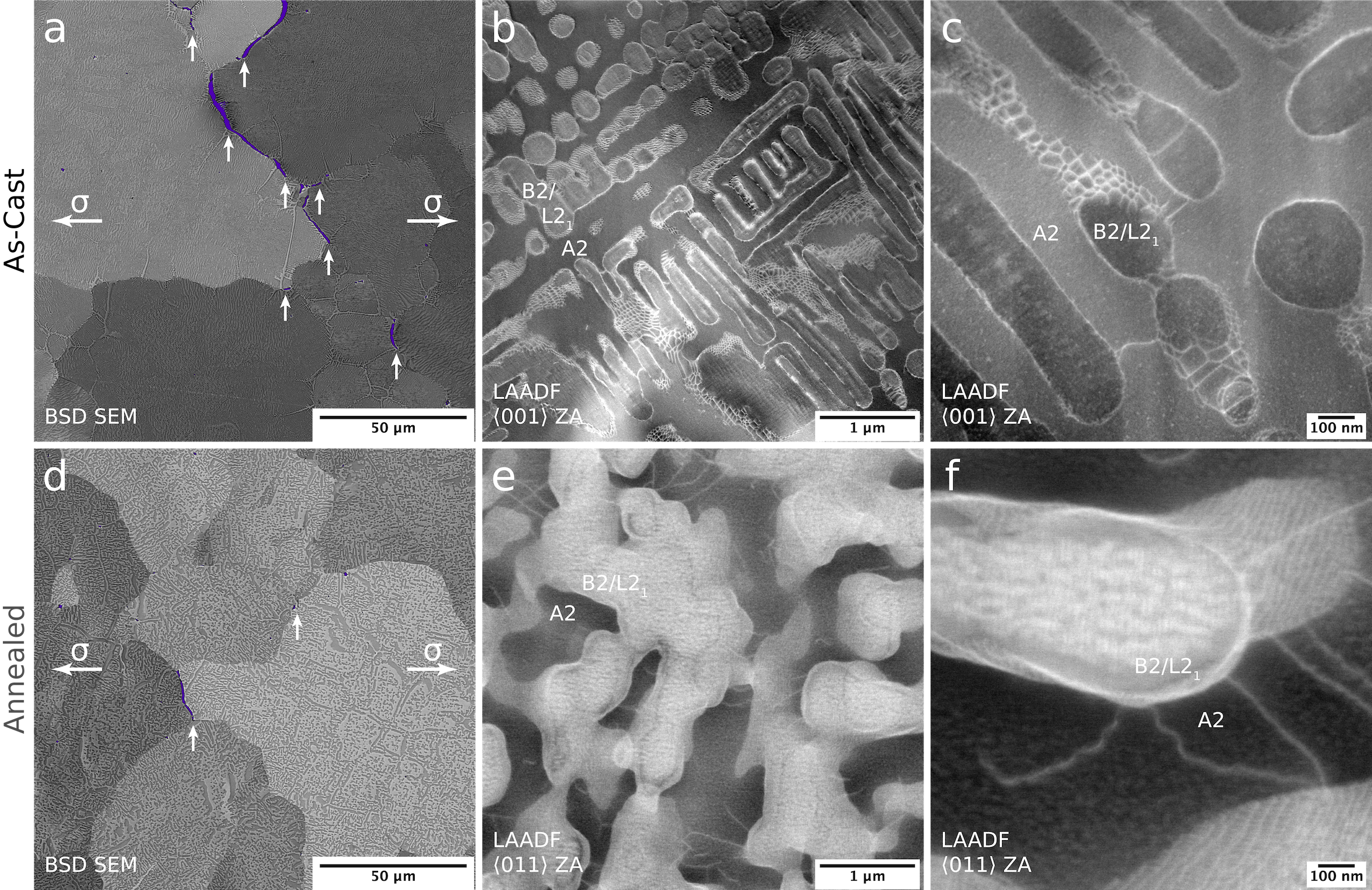}
\caption{\label{fig:d2_crept} Microstructures of as-cast and annealed alloy D2 after creep to 1.3\% and 2.0\% strain, respectively, at 800 $^{\circ}$C and 200 MPa. (a) and (c) show BSE SEM images of the grain structure, with creep cavities highlighted in purple and indicated with white arrows. (b) shows an overview LAADF image of the crept as-cast material, while (c) shows a higher magnification HAADF image. Both (e) and (f) show HAADF images of the crept annealed material, with the detector set to a longer camera length than in (c) to show specifically strain contrast.}
\end{figure}

For as-cast and annealed alloy D2, we interrupted a creep test conducted at 800 $^{\circ}$C under 200 MPa loading after the samples had reached the steady state creep regime. For the as-cast sample, we interrupted the test at about 1.3\% strain, and for the annealed sample, at about 2.0\% strain, at which point the sample had just begun to show an increase in strain rate indicative of the onset of tertiary stage creep. We conducted SEM observations on the surfaces parallel to the stress axis after polishing. Regions of these surfaces are shown in~\cref{fig:d2_crept}(a) and (d), where the stress axis has been indicated for both.\\

For the as-cast crept sample, a large number of thin cavities are observed throughout the gauge length, almost all located at grain boundaries which lie perpendicular to the stress axis. Other than the grain boundary cavities, there are no noticeable changes to the alloy microstructure after creep to 1.3\% strain in the SEM. Observations at higher magnification were conducted on a thin foil of the crept material using scanning TEM.~\cref{fig:d2_crept}(b) shows a diffraction contrast image of a representative region of the sample. We observe very slight coarsening of the microstructure compared to the untested as-cast material, but only a marginally greater density of mobile dislocations, where the majority of these mobile dislocations are seen to be located in the maze-like interdendritic regions of the material, with very few seen in the matrix of the dendritic regions. This observation suggests that dislocations are able to move much easier through the dendritic regions of the sample and tend to collect in the interdendritic regions. We can also see from the overview image that some of the mobile dislocations appear to be moving through the ordered phase domains as well as through the disordered phase. When compared to the LAADF image of the untested as-cast D2 specimen, shown in~\cref{fig:STEM}(e), we see that the number of mobile dislocations in the disordered phase of the crept specimen is only slightly greater than in the virgin material, while in the ordered phase, dislocations appear only after creep testing. This is somewhat surprising given that dislocation motion in the B2 and L2$_1$ phases should have a significantly higher energy barrier than in the A2 phase. This behavior could be a result to the very high stress level that the specimen was crept at, whereby dislocations in the disordered phase are able to shear through the ordered phase and/or nucleation sources in the ordered phase become activated. We take a closer look at the structure in~\cref{fig:d2_crept}(c) and observe clear indications of at least one dislocation in the ordered phase.\\

For annealed alloy D2, we also observe several grain boundary cavities in the crept state, however, these cavities are observed to occupy roughly half of the area that they occupy in the as-cast sample, despite the annealed alloy having been crept to a larger total strain level, in the tertiary regime. The HAADF image of~\cref{fig:d2_crept}(e) shows a higher number of mobile dislocations in the disordered phase domains than seen in the as-cast crept sample. Although not shown here, we observe significant number of dislocations in the ordered phase domains. We also observe the presence of large numbers of dislocation loops in the ordered phase, which may have formed by Orowan looping around disordered particles in the ordered phase. The interfacial dislocation networks in the annealed specimen appear to have a finer spacing than in the as-cast specimen, which could be related to the observation of a greater lattice parameter mismatch in the annealed alloy than the as-cast alloy in the untested condition, but could also arise from mobile dislocations being getting trapped at the interfaces and being incorporated into the networks.\\ 

We observe a large degree of secondary precipitation in both phases in the crept annealed specimen. Although not clearly seen in the HAADF image shown in~\cref{fig:d2_crept}(f), the secondary precipitates in the ordered phase were found to be finer, more numerous, and closer spaced than in the as-cast specimen, shown in (c). Importantly, we note that the disordered phase in the annealed condition contains a large number of very fine (5-10 nm diameter) secondary B2-ordered precipitates, while in the as-cast condition, no secondary precipitation is observed in the disordered phase. This suggests that these fine secondary domains may have formed during the annealing heat treatment, or upon cooling from same. The dislocations in the A2 phase seen in (f) are very wavy, suggesting multiple pinning points were present and prevented free dislocation motion during creep deformation. These pinning points are likely to correspond to the secondary precipitates, although this has not been determined definitively.\\

For both the as-cast and annealed specimens, we observe strong indications that the specimens deformed to some degree by enhanced diffusion, especially for the as-cast specimen. For the annealed specimen we also observe deformation to occur by dislocation climb and glide in both ordered and disordered phases. These observations align well with the observed creep exponents obtained from the steady state creep curves, with the as-cast condition showing a creep exponent which indicates mostly diffusion-based deformation, and the annealed condition, having a creep exponent which is better explained by dislocation glide and climb processes.\\


\section{Discussion}

Several microstructural features, at differing length scales, affect the overall high temperature mechanical response of these three alloys. At the lowest length scale is the composition of individual ordered and disordered domains, and the effect that the composition has on phase stability, diffusion rate, lattice parameter mismatch, and elastic response. The next microstructural feature that varies across the three alloys is the morphological relationship between ordered and disordered domains, with regions of equiaxed ordered precipitates in a disordered matrix (matrix-precipitate morphology) and regions of interpenetrating maze-like ordered and disordered domains (maze-like morphology). The relative phase fraction, precipitate size, domain width, and interfacial coherency are all expected to strongly effect the properties of both morphology types observed. Within individual grains, the ratio of matrix-precipitate to maze-like morphology will influence the mechanical response, as will the phase character and composition at grain boundaries. In the following sections, we present a discussion of the various strengthening contributions across these three alloys, a comparison of mechanical properties with other recently developed HEA/CCAs intended for similar applications, and finally, a more detailed analysis of the creep behavior in the D2 alloy.\\

\subsection{Strengthening Contributions}

We will discuss the different possible strengthening contributions in these three alloys in the following order: $(i)$ lattice friction resistance in terms of the Peierls barrier, $(ii)$ Orowan particle strengthening, $(iii)$ solid solution strengthening, $(iv)$ dislocation shearing of ordered phase domains. As all three alloys have similar grain sizes in both as-cast and annealed conditions, the average grain size is not expected to contribute to differences in observed strength and ductility.\\

\subsubsection{Lattice Friction Resistance}

The first contribution to yield stress to consider is the Peierls flow stress. We will carry out our analysis with the assumption that the first phase to yield will be the disordered A2 phase. With that said, however, Schroll et al. have compiled results from several studies of B2-NiAl which make use of embedded atom method (EAM) potentials to calculate values of the Peierls stress for different slip systems and have found values as low as 60 MPa (edge-type) and 70 MPa (screw-type) for the 
``soft" $\langle100\rangle\{01\overline{1}\}$ slip system in NiAl~\cite{Schroll1998, Ludwig1995}. These low values for Peierls stress only occur for certain orientations, but may play a role during deformation of the studied alloys. Very little data is available on the Peierls barrier in L2$_1$-Ni$_2$TiAl and -Fe$_2$TiAl, but given the extra degree of superlattice ordering in this phase, the operative slip system ($\langle011\rangle\{01\overline{1}\}$)~\cite{Strutt1976} has a greater Burgers vector length than that in either A2 or B2 phases, and the resulting Peierls barrier will be significantly higher.\\

Although there is great discrepancy in the literature as to the exact formulation of the Peierls stress, one common feature between the analytical expressions that have been developed is a linear dependence on the isotropic shear modulus (we will not treat anisotropy in our analysis). The true flow stress contribution from lattice friction will be lower than the Peierls stress, especially at high temperatures, due to thermal activation of dislocation motion, in particular due to the kink-pair mechanism of dislocation glide~\cite{Lothe1959, Hull2001}. For a given temperature, however, the lattice friction stress in individual disordered phase domains will vary across these three alloys due only to differences in the shear modulus induced by disordered phase composition. Based on data collected from several sources, Ghosh and Olson have compiled values for the change in $\alpha$-Fe shear modulus, $G$, with composition, $c_x$, for a large number of atomic species, $x$~\cite{Ghosh2002}. From the $dG/dc_x$ values listed in that work, we can relate the change in shear modulus of the disordered phase due to composition according to,\\
\begin{equation}
    \Delta G(c) = -100.85c_{Al}+34.06c_{Cr}-90.65c_{Ni}-74.59c_{Ti}.
\end{equation}
We assume here that there are no non-linear effects due to interaction of solute species and that the chemical dependence of the shear modulus does not change with temperature. For the compositions shown in~\cref{fig:STEM}, the calculated increase in shear modulus of the A2 phase is about 14.2 GPa in alloy D1, 8.6 GPa in alloy D2, and 4.5 GPa in alloy D3. In order to see how the shear modulus values of the three alloys compare at high temperature, we also have to consider the shear modulus temperature dependence. For pure polycrystalline $\alpha$-Fe, this has been fit empirically by~\cite{Ghosh2002},\\
 \begin{equation}
     G(T)=8.407\left[1-0.48797\left(\frac{T}{T_C}\right)^2+0.12651\left(\frac{T}{T_C}\right)^3\right]\times10^{10}~\mathrm{Pa},
 \end{equation}
where $T_C$ is the Curie temperature (1043 K for pure $\alpha$-Fe). The value of the shear modulus determined using the above equation ranges from 57 GPa at 700 $^{\circ}$C to 47 GPa at 900 $^{\circ}$C. Combining these values with the composition dependence, the respective D1, D2, and D3 shear moduli are 71.3, 65.6, and 61.5 GPa at 700 $^{\circ}$C, and 61.6, 55.9, and 51.8 GPa at 900 $^{\circ}$C. Although the Al, Ni, and Ti contents in the A2 phase are likely to be higher---and the shear modulus increase correspondingly lower---than the values determined, the calculated values indicate that the lattice friction resistance to dislocation motion is greatest in alloy D1 and least in alloy D3. However, given the observed variation in yield strength across the three alloys, with alloy D1 having the lowest yield strength, and alloy D3 having the highest yield strength, other effects are expected to play a much larger role than the magnitude of the Peierls stress.\\

\subsubsection{Orowan Strengthening}

\begin{figure}[h!]
\centering
\captionsetup{skip=4pt}
\includegraphics[width=0.95\textwidth]{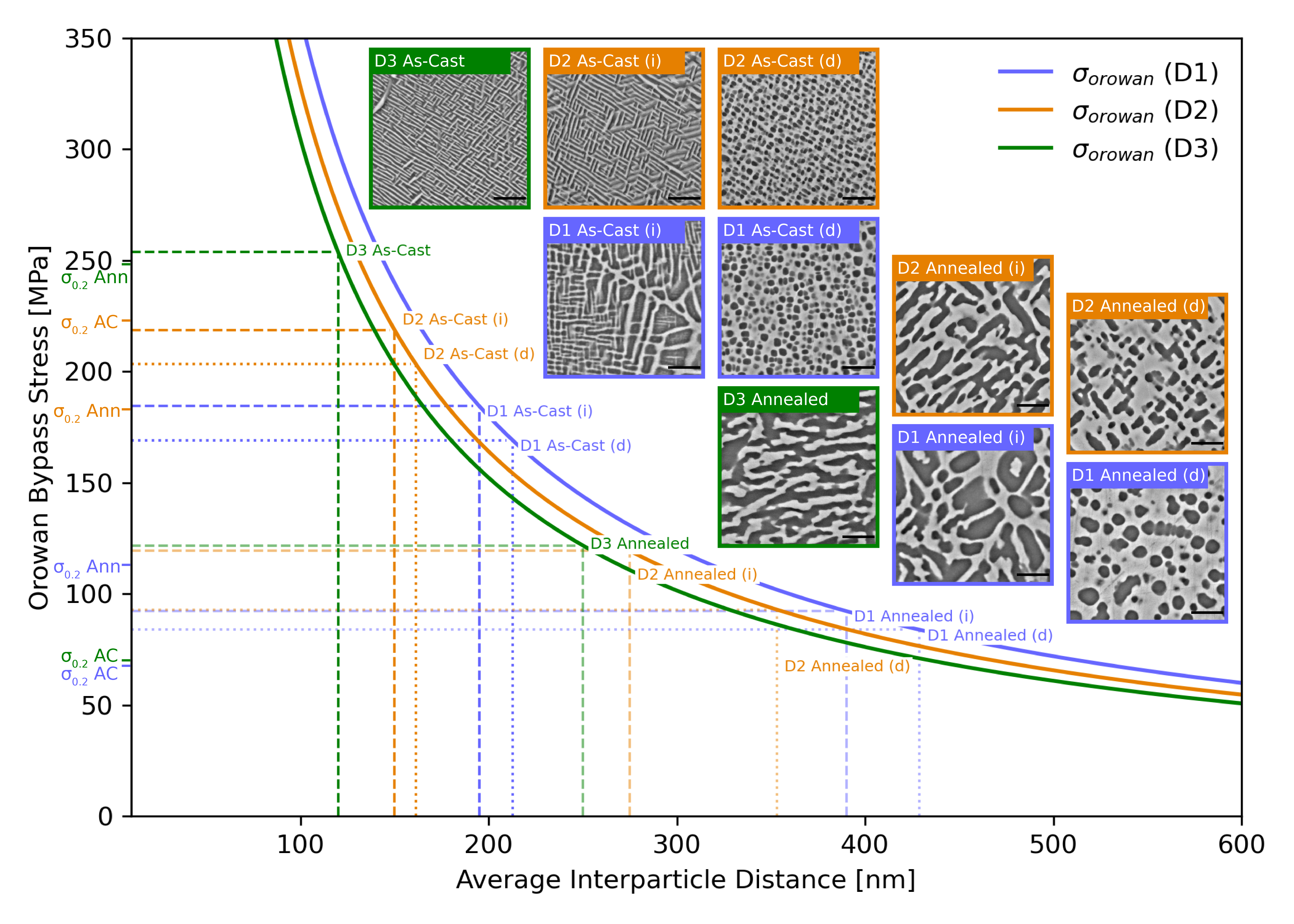}
\caption{\label{fig:orowan}Orowan bypass stress in the A2 phase versus average interparticle distance in all morphology types present in alloys D1-D3. Shear modulus values for the A2 phase are calculated at 900 $^{\circ}$C. Scale bar in images is 2 $\mu$m.}
\end{figure}

We will consider the general Orowan bypass mechanism with the form~\cite{Abe2008},\\
\begin{equation}
    \sigma_{orowan}=0.81M\frac{Gb}{\lambda},
\end{equation}
where $M=2.9$ is the Taylor factor for randomly oriented bcc polycrystals~\cite{Kocks1970}, $G$ is the temperature and composition adjusted shear modulus of the A2 phase, $b$ is the Burgers vector length, and $\lambda$ is the average ordered phase domain separation within which dislocations will glide in the disordered phase, which we will refer to as the interparticle distance. For the maze-like domains, we take the average interparticle distance to be simply the average A2 domain width, values of which have been determined from the BSE SEM images of the alloy microstructures. This is similar to the method that is used to estimate the Orowan stress in Ni-base superalloys with high $\gamma'$-Ni$_3$Al phase content, where the interparticle distance is taken as the average disordered phase channel width~\cite{Fedelich2009}. For the dendritic domains in alloys D1 and D2, the interparticle distance can be calculated from the average particle radius, $\langle r\rangle$, and volume fraction of precipitates, $f$, according to~\cite{Song2017a},\\
\begin{equation}
    \lambda=\langle r\rangle*\left[\sqrt{\frac{2}{3}}\left(\sqrt{\frac{\pi}{f}}-2\right)\right].
\end{equation}
With values for $\lambda$ determined for both morphology types in all three alloys, we can compare the Orowan bypass stress expected for each of the individual regions in both as-cast and annealed conditions. We show this graphically in~\cref{fig:orowan}, where the Orowan stress curves have been calculated using values of the shear modulus of the A2 phase at 900 $^{\circ}$C, taking into account contributions from solute species determined earlier. We observe that although alloy D1 has a greater Orowan stress than the other two alloys for a given interparticle distance, the greater values of interparticle distance determined for the alloy lead to the lowest total Orowan stresses, in both as-cast and annealed conditions. For alloys D1 and D2 in the as-cast state, the average interparticle distance in the interdendritic regions is lower than in the dendritic regions, indicating that the maze-like structure in the interdendritic regions will have a higher yield strength than the matrix-precipitate structure. However, the difference in strengthening is not very large considering the difference in ordered phase volume fraction in the respective regions. In the annealed condition, we see a larger spread in average interparticle distances, especially between the dendritic and interdendritic regions of alloys D1 and D2, but the overall Orowan strengthening contributions show little spread, all lying between about 80 and 120 MPa.\\

We have included the 900 $^{\circ}$C 0.2\% offset yield stress values listed in~\cref{tbl:tension} on the stress axis in~\cref{fig:orowan}. Comparing the values for Orowan bypass stress with the determined yield stress we note several interesting trends. In the as-cast condition, the Orowan stress for alloys D1 and D3 significantly overestimates the yield stress, presumably due to premature failure of the material. The pre-yield failure of the as-cast D3 material in all tested samples indicates that the cohesion of the grain boundaries was too low to allow even microplasticity, and/or that the maze-like structure does not allow for continuous glide in the disordered phase channels. As alloy D2 was also composed of predominantly the maze-like microstructure, yet was consistently able to reach yielding, we conclude that the presence of the dendritic regions, with their characteristic disordered matrix, allowed a smooth increase in stress in these regions up to the point of yielding. This has been observed in complementary FEM simulations and neutron diffraction experiments for bulk materials with the same morphology as in the dendritic regions~\cite{Song2015}. In as-cast alloy D1 we would have expected the samples to reach yielding as well, as these samples had an even higher portion of the matrix-precipitate morphology than the D2 alloy, but we were unable to obtain a sample which showed true yielding behavior. This is likely due to the large fraction and size of ordered phase domains near the grain boundaries, as well as high Ti segregation to the grain boundaries, as seen in~\cref{fig:sem_eds}. It is also possible that the presence of fine secondary B2 precipitates in the disordered phase, as observed in~\cref{fig:STEM}(d), decreased the effective interparticle distance in the dendritic regions, pushing the required Orowan bypass stress in these regions higher. For as-cast alloy D2, where we observed almost 3\% plastic strain before failure, we see that the measured yield stress (223 MPa) is slightly larger than the determined Orowan stress in the dendritic regions (203 MPa) and is nearly the same as that in the interdendritic regions (219 MPa). However, given the observation that the maze-like microstructure was unable to reach yielding for alloys D1 and D3, we assume that the Orowan stress in the dendritic regions is closer to the real contribution to yield stress due to the Orowan mechanism. We note that the yield stress determined for as-cast alloy D2 is likely a lower bound for the material, given the high degree of intergranular fracture observed throughout the fracture surface; the appearance of yielding behavior may have corresponded with stable micro-crack growth, rather than true plasticity. It remains to be determined how the stress and strain evolve in maze-like regions in all three alloys, and indeed, how the macroscopic yield stress and subsequent hardening behavior of alloys D1 and D2 result from the interaction of these maze-like domains with the dendritic matrix-precipitate type regions.\\

The calculated values of Orowan stress for the annealed samples in all cases underestimate the observed yield stress values. For alloy D3 the calculated Orowan stress (122 MPa) severely underestimates the 0.2\% offset yield stress (248 MPa), which indicates that the Orowan formulation may not accurately describe the strengthening effect of the maze-like microstructure. The observed yield stress in annealed alloy D2 (183 MPa) is also significantly higher than the calculated Orowan strengthening contribution from either interdendritic regions (117 MPa) or dendritic regions (93 MPa). Here, also, we note that the strengthening contribution from interdendritic regions may be much higher than the calculated value. A superposition of strengthening contributions from dendritic and interdendritic regions may thereby result in an overall bypass stress which is much closer to the observed value. The high yield stress values in the annealed alloys may also be partially explained by the presence of secondary precipitates in the disordered phase domains, clearly seen for annealed alloy D2 in~\cref{fig:d2_ann_STEM}, which decrease the effective interparticle distance. Whether these secondary precipitates remain stable and are thereby able to contribute to a higher required bypass stress at high temperatures is unclear, but depends on the solubility of the constituent elements (mostly Ni and Al) at the testing temperature. For annealed alloy D1 the observed yield stress (113 MPa) is much closer to the Orowan strengthening contributions (92 MPa interdendritic and 84 MPa dendritic), however, we note from the low plastic strain to failure for this alloy (0.5\%) that the determined yield stress value may be a lower bound due to stable growth of grain boundary cracks, rather than true yielding behavior.\\ 

\subsubsection{Solid Solution Strengthening}

The A2 phase in the three alloys studied contains large concentrations of substitutional solute species, particularly Al and Cr, as seen in~\cref{fig:STEM}. The resulting supersatured solid solution is strengthened by distortions in the lattice caused by differential volumetric and shear modulus effects of the solute species, which act as additional barriers to dislocation motion~\cite{Fleischer1963, Nabarro1977, Labusch1970}.  Takeuchi studied the solid solution strengthening effect of several solute species in $\alpha$-Fe and found that the influence of increasing solute concentration on modulus normalized critical resolved shear stress (CRSS) follows a linear trend. For solutions containing up to 5.85 at.\% Al, the change in modulus normalized critical resolved shear stress with Al concentration has been found to be $d(\tau_{ss}/G)/dc_{Al}=0.014$~\cite{Takeuchi1969}. In a similar fashion the effect of Cr has been found to be $d(\tau_{ss}/G)/dc_{Cr}=0.0074$~\cite{Horne1963}. For the calculation of solid solution strengthening contribution in these three alloys we then have,\\
\begin{equation}
    \Delta\sigma_{ss}=0.81MG_c\left[0.014c_{Al}+0.0074c_{Cr}\right],
\end{equation}
where $M=2.9$ is the Taylor factor for randomly oriented bcc polycrystals~\cite{Kocks1970}, and $G_c$ is the shear modulus of the composition in question. The respective D1, D2, and D3 solid solution strengthening contributions are then 581, 563, and 505 MPa at 700 $^{\circ}$C, and 502, 480, and 425 MPa at 900 $^{\circ}$C. These values are in all cases much larger than the experimentally observed yield stress values, but are closer in magnitude for the 700 $^{\circ}$C results. This is easily explained by thermal effects which reduce the effective free energy barrier that prevents movement of dislocations past the substitutional obstacles as temperature is increased. This strengthening contribution, along with the Peierls barrier, are expected to play a larger role at lower temperatures, where less thermal energy is available to overcome these barriers. These lattice friction contributions may be the reason that the yield stresses observed at 700 $^{\circ}$C are significantly higher than the Orowan bypass stresses present at this temperature, which are only slightly larger than those calculated at 900 $^{\circ}$C due to the increase in shear modulus with decreasing temperature.\\

The relatively weak energetic barrier presented by homogeneously distributed substitutional solute species may be substantially increased due to migration of the solutes to defect cores, a phenomenon that results in so-called Cottrell atmospheres~\cite{Cottrell1949, Hull2001}. This leads to what is known as dislocation locking, whereby the stress required for dislocation motion is increased as a result of the increased interaction energy of the dislocation with the solute atmosphere. This effect, if indeed active in these alloys, would be expected to impact the performance of the annealed material greater than the as-cast material since the long holding time at 900 $^{\circ}$C allows relief of the thermal stresses present from casting, and subsequent diffusion of the solute species to the dislocations formed during the stress relief. Therefore solute atmosphere locking is another possible mechanism that could explain the discrepancy between the Orowan bypass strengthening contribution and the observed yield strength for the annealed alloys. We point out, however, that we do not observe indications of a yield drop which is characteristic for strong dislocation locking in the stress strain curves shown in~\cref{fig:tension}.\\

\subsubsection{Dislocation Shearing of Ordered Phase Domains}

A past work on Fe-base superalloys strengthened by B2 and L2$_1$ precipitates calculated that the precipitate shearing mechanism requires stresses on the order of 1 GPa, when the volume fraction of B2 precipitates was only 16\%~\cite{Vo2014}. This calculation took into account stress arising only from the creation of new APB surfaces, neglecting contributions from coherency strains and shear modulus mismatch. As all alloys studied here have much higher ordered phase volume fractions, the precipitate shearing stress arising from just the creation of APB surfaces will be even larger than the value found be Vo et al. We can thereby conclude that shearing is not likely to dictate the yielding behavior in any of the morphologies and  material conditions observed, without needing to account for the other contributions to the ordered phase shearing strengthening mechanism. The only case where this may not hold true is for nanoscale secondary precipitates which may be present in some material conditions, where the small average radii may allow dislocation shearing to occur at stresses lower than those required for Orowan bowing.\\


\subsection{High Temperature Specific Yield Stress Comparison} 

\begin{figure}[h!]
\centering
\captionsetup{skip=4pt}
\includegraphics[width=0.65\textwidth]{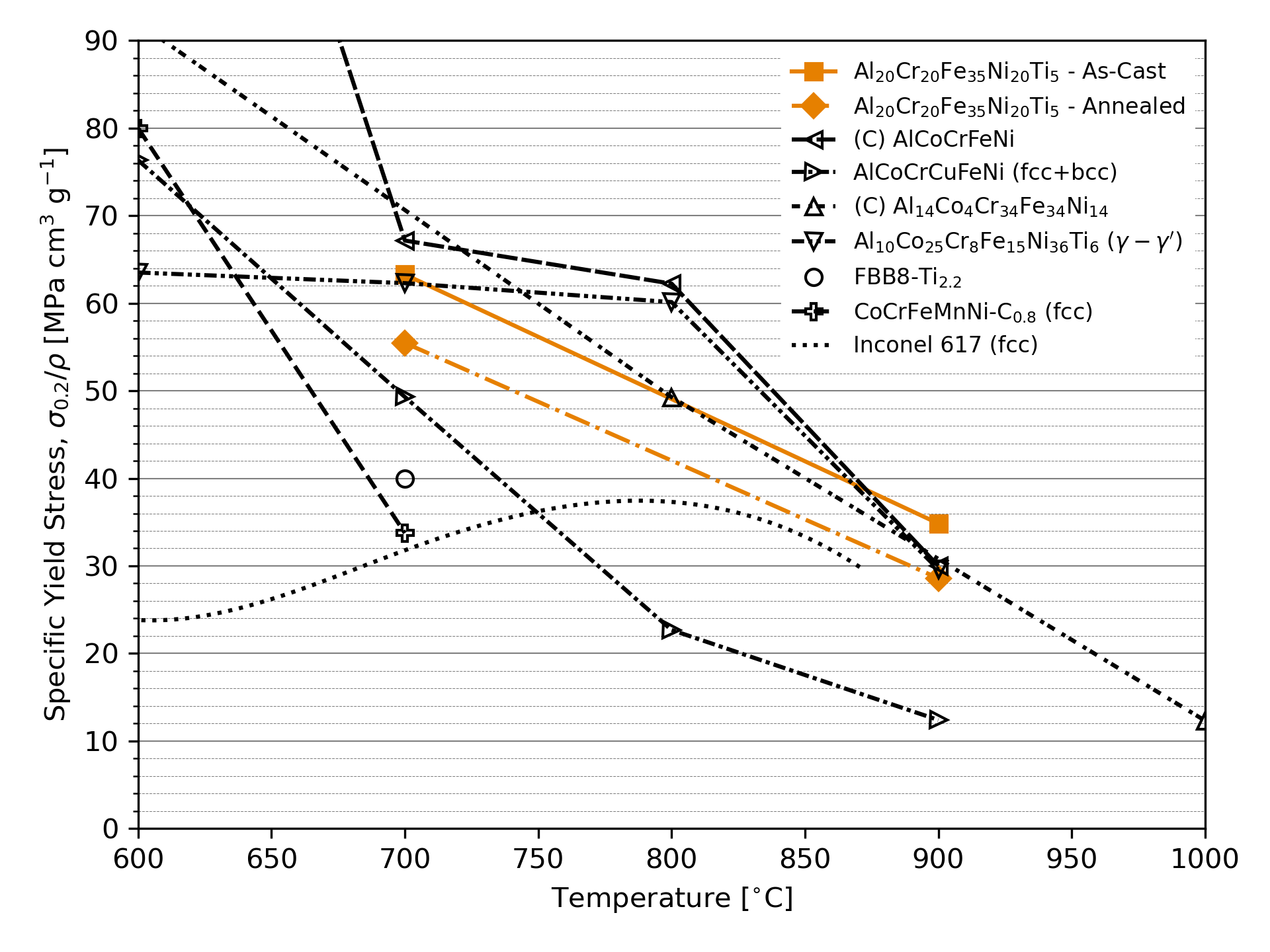}
\caption{\label{fig:syscomp}Specific yield stress versus temperature for alloy D2 (\Two) as well as several other HEA/CCAs designed for high temperature application, an optimized Fe-base superalloy, and Inconel 617. After Refs~\cite{Lim2017, Kuznetsov2012, Zhou2018, Haas2019, Song2015, Lu2020, McCoy1985}.}
\end{figure}

For high temperature applications involving rotating parts, where primary stresses arise due to the weight of the part itself, the specific tensile yield stress of the material is one of the main performance determining properties. In this sense it is valuable to see how the specific yield stress (SYS) of the D2 alloy compares to other high temperature oriented HEA/CCAs and conventional alloys in the temperature range of interest. We have included such a comparison in~\cref{fig:syscomp} between 600 and 1000 $^{\circ}$C. We see that alloy D2 has similar SYS to equiatomic AlCoCrFeNi
~\cite{Lim2017} and Al$_{14}$Co$_{4}$Cr$_{34}$Fe$_{34}$Ni$_{14}$~\cite{Zhou2018} CCAs, which were both tested in compression. It is likely, however, that both of these alloys will show lower SYS in tension. The AlCoCrCuFeNi alloy, which contains both bcc and fcc phases, shows a lower SYS than alloy D2 between 700 and 900 $^{\circ}$C~\cite{Kuznetsov2012}. An interstitially strengthened HEA (iHEA) based on the Cantor alloy shows high SYS at 600  $^{\circ}$C, but has lost much of this strength at 700 $^{\circ}$C, showing a SYS roughly half that of as-cast alloy D2
~\cite{Lu2020}. The ferritic superalloy, FBB8, containing a bcc matrix with hierarchical L2$_1$+B2 precipitates by incorporation of 2.2 at.\% Ti~\cite{Song2015}, shows a lower SYS at 700 $^{\circ}$C than both as-cast and annealed alloy D2. Perhaps most notably, the SYS of alloy D2 competes directly with polycrystalline $\gamma-\gamma'$ structured Al$_{10}$Co$_{25}$Cr$_{8}$Fe$_{15}$Ni$_{36}$Ti$_{6}$, even showing a higher SYS at 900 $^{\circ}$C in the as-cast state, while outperforming commercial Inconel 617.\\

\subsection{Creep Behavior of \Two}

\begin{figure}[h!]
\centering
\captionsetup{skip=4pt}
\includegraphics[width=0.95\textwidth]{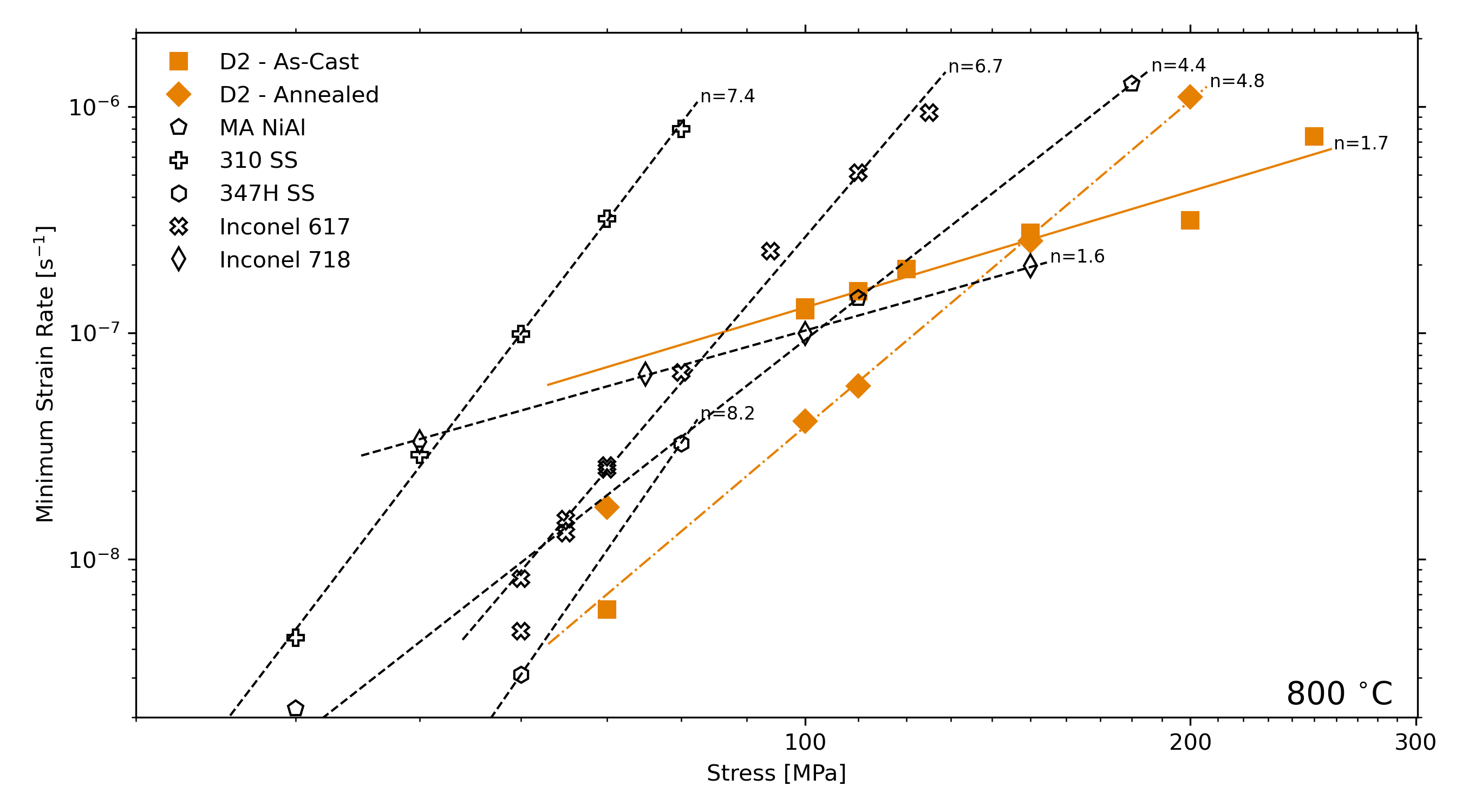}
\caption{\label{fig:creepcomp}Minimum steady-state creep rate plot for alloy D2 tested at 800 $^{\circ}$C. Minimum rates for several other high temperature alloys tested at 800$^{\circ}$C are included for comparison. After Refs~\cite{Amirkhiz2015, Amirkhiz2019, Benz2014, Chen2017, Suh1995}.}
\end{figure}

We show the 800 $^{\circ}$C steady-state creep rate values for alloy D2 in comparison with two high temperature stainless steels~\cite{Amirkhiz2015, Amirkhiz2019}, two common Ni-base superalloys~\cite{Benz2014, Chen2019}, and a high performing mechanically alloyed (MA) NiAl alloy~\cite{Suh1995} in~\cref{fig:creepcomp}. We see that alloy D2 performs very well in comparison to these other alloys at this temperature. In the as-cast condition, alloy D2 has only slightly higher strain rates than Inconel 718, and more interestingly, has nearly the same power law exponent for stresses above 100 MPa. At this temperature, Inconel 718 has been observed to fail primarily by the nucleation and linkage of voids in the structure~\cite{Chen2019}, similar to what was observed in the as-cast D2 material. This suggests that this process, which is mainly controlled by diffusion and grain boundary cohesive strength, is responsible for the low value of the creep exponent in as-cast alloy D2 at 800 $^{\circ}$C and above 100 MPa. Furthermore, it indicates that the as-cast structure has sufficient resistance to dislocation glide and climb up to 200 MPa to cause the weakest part of the structure, namely the grain boundaries, to fail. If the grain boundaries can be strengthened, or eliminated altogether by producing a single crystal of the material, the creep strain rates at this temperature may be significantly lower. The creep rates in the as-cast structure are therefore a hard upper bound to the true creep behavior of this material. When the alloy was tested under 70 MPa loading, we observe a marked decrease in strain rate, which suggests the presence of a threshold creep stress, as has been observed to be in a similar range in other precipitation strengthened alloys, notably several Fe-base superalloys tested at 700 $^{\circ}$C~\cite{Vo2014, Song2015, Rawlings2017}. For those studies, it was found that the threshold stress corresponds to the stress necessary for general climb inhibited by elastic stresses arising from lattice parameter mismatch between the matrix and precipitate phases. In the case of as-cast alloy D2, however, stresses above about 70 MPa at 800 $^{\circ}$C lead to increasing rates of cavitation type creep, which is controlled by a combination of bulk diffusion and dislocation movement. Therefore, the underlying reason for the presence of threshold type behavior must be more critically evaluated for this complex alloy.\\

In analysing the creep behavior of as-cast alloy D2, one cannot neglect the influence of the distinct dendritic and interdendritic morphologies. We saw from our analysis of the Orowan stresses which arise from the different sample regions that the dendritic regions should reach the yield point before the interdendritic regions. In analogy, deformation processes which occur at lower stress than dislocation glide, in particular dislocation climb, are also expected to occur first in the dendritic regions of the sample. In our observations of the as-cast D2 sample crept to 1.3\% strain at 800 $^{\circ}$C, 200 MPa, we found a far lower density of mobile dislocations in the dendritic portion of the sample than in the interdendritic portion. If this observation is assumed to be representative of the general behavior within the material, this suggests that dislocations are able to move relatively easily through the dendritic pockets within the material but are comparatively trapped within the interdendritic maze-like regions. The presence of such inhomogeneous yielding behavior could lead to stress concentrations within the interdendritic regions, with the ultimate outcome being grain boundary decohesion, as the grain boundaries are nearly invariably located at the intersection of adjacent interdendritic regions. Below a certain stress, dislocation motion, even by general climb, may not be possible within the dendritic regions of the material, with the result that stress concentrations at grain boundaries, and resulting grain boundary decohesion, would no longer be present.\\
 
When we compare the power law fit for MA NiAl with that of annealed alloy D2, we note a very similar creep exponent and slightly lower strain rates in alloy D2. This suggests that the creep deformation in annealed alloy D2 may also occur in large part in the ordered phase domains, rather than just in the disordered phase. We must qualify this observation by noting that the creep tests in MA NiAl were performed in compression rather than in tension. Still, this finding is in accordance with our observations of mobile dislocations in the ordered phase after creep in annealed alloy D2. The observed power law exponent of approximately 4.8 is suggestive of a combined glide and climb creep mechanism, however, these processes may be occurring simultaneously in both the ordered and disordered phases.\\   

\subsubsection{Diffusion Rates}

At the high temperatures and low strain rates encountered during creep, the diffusion rate in the B2 and L2$_1$ phases plays a significant role in determining the steady-state creep rate as well as the rate of coarsening. Rudy and Sauthoff studied the creep behavior in B2 structured (Fe, Ni)Al alloys and found that the maximum creep resistance in these alloys corresponded to the alloy with the minimum interdiffusion coefficient~\cite{Rudy1986}. The minimum interdiffusion coefficient was found to occur for stoichiometric Al occupation and an Fe:Ni ratio of about 1:4, corresponding, respectively to a maximum in long-range order parameter, and maximum (Fe, Ni) sublattice disorder. The significance of these findings for the ordered phase domains in the alloys studied here is that alloy D1 is observed to contain nearly an ideal Fe:Ni ratio, especially in the dendritic regions, but is also expected to have the greatest number of antisite defects, as indicated by the low Al and Ti occupation. Alloy D2, which has a slightly greater Fe:Ni ratio, but a significantly higher Al and Ti occupation, is expected to have an overall lower set of ordered phase interdiffusion coefficients than alloy D1. Alloy D3 has nearly the same Al and Ti occupation as alloy D2, but has a higher than optimal Fe:Ni ratio, indicating that the interdiffusion coefficients in this alloy are also higher than in alloy D2. These findings suggest that alloy D2 may have the lowest overall diffusion based creep rate of the three alloys studied, but could be further optimized by slightly reducing the Fe:Ni ratio, and increasing the Al and Ti sublattice occupation.\\

The diffusion rate in the disordered phase domains of these three alloys may play an even larger role on the creep response than the diffusion rate in the ordered phase, due to climb of mobile dislocations in the disordered phase past ordered phase obstacles, at stresses far below the yield stress. Although a quantitative determination of the interdiffusion coefficients in the A2 phase is beyond the scope of this work, we can make some general remarks about the trends that can be expected. From Darken's second equation
~\cite{Darken1948} one can infer that increasing concentration of an element which has a greater solute diffusion coefficient in a given phase than the self-diffusion coefficient of the host species will increase the interdiffusion coefficient of the mixture relative to the pure species. For $\alpha$-Fe (in the paramagnetic state) the ratio of the solute diffusion coefficients of Al, Cr, Ni, and Ti to the self-diffusion coefficient of Fe have been estimated as $D_{Al}/D_{Fe}=1.4$, $D_{Cr}/D_{Fe}=1.4$, $D_{Ni}/D_{Fe}=1.0$, and $D_{Ti}/D_{Fe}=3.9$~\cite{Oikawa1982}. From these values, we see that Ti should have by far the largest effect on the A2 interdiffusion coefficient in these alloys, Al and Cr should have a much lower effect, while Ni has little to no effect. It is tempting to draw firm conclusions from these values, however, it is unclear  over which composition ranges these relationships hold. It is possible, for example, that when the disordered phase approaches equal parts Fe and Cr, as is the case for alloys D1 and D2, the interdiffusion coefficient may show anomalous behavior. Furthermore, magnetic interactions in $\alpha$-Fe have been shown to play a very large role in diffusional processes~\cite{Bhadeshia2001}. With these reservations in mind, we reason that in all three alloys, the disordered phase interdiffusion coefficient is governed by the presence of a trace Ti content, and will be approximately the same for all three alloys within the accuracy of our composition measurements.\\

\section{Conclusions}

Our characterisation of alloys D1 (Al$_{16}$Cr$_{25}$Fe$_{35}$Ni$_{20}$Ti$_4$), D2 (Al$_{20}$Cr$_{20}$Fe$_{35}$Ni$_{20}$Ti$_5$), and D3 (Al$_{24}$Cr$_{15}$-\\Fe$_{35}$Ni$_{20}$Ti$_6$) reveals a clear gradient in several microstructural features that closely follows the compositional gradient across the alloys. The first length scale at which differences become noticeable is at the scale of individual grains, where the composition difference between dendritic and interdendritic areas is greatest for alloy D1 and least for alloy D3, as demonstrated in~\cref{fig:sem_eds}. For alloys D1 and D2, the large degree of dendritic segregation results in noticeable phase fraction and morphology differences between dendritic and interdendritic regions. Due to the high volume fraction of both ordered and disordered phases, it is not reasonable to associate these alloys most closely with either ferritic superalloys or aluminide/heusler phase materials. For the composition range studied here, the resulting alloys would best be described as ferrite-aluminide ($\alpha-\beta$) duplex alloys. For these duplex alloys we have thoroughly characterized the as-cast and annealed structures, as well as the high temperature tensile and creep behavior. We make the following conclusions:\\

\begin{itemize}
    \item By varying the ratio of (Al+Ti):Cr, we are able to tune the degree of dendritic segregation in these alloys.
    \begin{itemize}
        \item On the high Cr end of the gradient, the majority of the structure is composed of primary dendrites with an A2 matrix and equiaxed B2/L2$_1$ precipitates.
        \item As Cr is replaced by Al and Ti, the volume of the primary dendrites is replaced by increasing amounts of interdendritic structure, which is composed of a fine maze-like arrangement of B2/L2$_1$ and A2 phases.
    \end{itemize}
    \item The average precipitate size within dendritic regions, as well as the average domain width in maze-like regions were both found to decrease with increasing Al and Ti content, likely due to a decrease in the respective ordering and decomposition temperatures.
    \item The coarsening rate in all three alloys was observed to be approximately equivalent after 100 h annealing at 900 $^{\circ}$C.
    \item Disordered phase lattice parameters in both as-cast and annealed conditions were found to slightly increase with increasing Al and Ti content and decreasing Cr content, while ordered phase lattice parameters showed the opposite behavior. Corresponding bulk averaged lattice parameter misfit was found to decrease as Al and Ti were substituted for Cr.
    \item Insights gained from calculations of the theoretical Orowan bypass stress in different regions of the three alloys were used to explain the results of high temperature tensile experiments:
    \begin{itemize} 
        \item Embrittlement in the interdendritic regions of D1 (\One) led this alloy to fracture prior to yielding in the as-cast state at both 700 and 900 $^{\circ}$C. In the annealed condition, yield strengths were higher than predicted by the Orowan strengthening mechanism, possibly due to either the presence of nanoscale secondary precipitation, or solute atmosphere locking.
        \item Fully lamellar alloy D3 (\Three) failed prior to yielding in the as-cast state at both testing temperatures and furthermore showed near-zero ductility in the annealed state indicating the absence of continuous pathways for dislocation glide in the maze-like microstructure.
        \item Duplex alloy D2 (\Two) showed a balance between moderate ductility and high yield strength at high temperature, showing greater ductility than alloy D1, with its large volume fraction of ductile disordered matrix, and only slightly lower yield strength than the fully maze-like alloy D3. The Orowan strengthening contribution for the dendritic regions was predicted to be lower than that in the interdendritic maze-like regions, indicating that yielding should occur first in the primary dendrites. Indeed, at 900 $^{\circ}$C, the observed as-cast yield strength was only slightly greater than the calculated Orowan strengthening contribution from the dendritic regions. 
    \end{itemize}
    \item Creep tests carried out on alloy D2 under a range of loading conditions at 700, 750, and 800 $^{\circ}$C revealed promising creep resistance in both the as-cast and annealed conditions. For tests carried out at 800 $^{\circ}$C, investigation of the crept microstructure revealed that deformation in the as-cast structure occurred mainly by grain boundary cavitation, resulting in diffusion controlled creep behavior. This indicates that observed creep rates are an upper bound limited by grain boundary cohesion. Power law exponents in the annealed state suggest dislocation glide and climb controlled creep behavior. Observation of the annealed structure after creep testing reveals mobile dislocations in both ordered and disordered domains, indicating that both phases must be taken into account in predicting strain rates in the annealed condition.
\end{itemize}

\section*{Acknowledgements}

This work was sponsored by the Deutsche Forschungsgemeinschaft (DFG) within the priority programme SPP 2006 CCA-HEA, grant no. CL 2133/4-1. Dennis Klapproth, Frank Rütters and Jürgen Wichert all deserve thanks for material synthesis and heat treatment. Tristan Wickfeld and Bernd Deuerling are thanked for tension and creep specimen preparation. The authors would like to thank Benjamin Breitbach for XRD measurements and Andreas Janssen for tension testing. The authors gratefully thank Dr. Frank Stein and Dr. Martin Palm for fruitful discussions, as well as Professor Gerhard Dehm for careful and insightful editing of the manuscript and for general support over the course of this study.

\bibliography{mybibfile}

\end{document}